\documentclass[floatfix,twocolumn,amsmath,amssymb,nofootinbib]{revtex4}
\usepackage{multirow,array}
\usepackage{graphicx,color,rotating}
\usepackage{amssymb, amsmath}
\usepackage{epstopdf}
\usepackage{subfigure}
\DeclareGraphicsExtensions{.pdf,.png,.gif,.jpg,.eps,.ps}

\begin{document}
	
%----------------------------------------------------------------------
\title{Acceleration feature points of unsteady shear flows}
%----------------------------------------------------------------------
\author{\sf Jens Kasten}
\affiliation{\small\sl Leipzig University, PF 100 920, 04009 Leipzig, Germany}
\email{kasten@informatik.uni-leipzig.de}

\author{\sf Jan Reininghaus}
\affiliation{\small\sl IST Austria, Am Campus 1, 3400 Klosterneuburg, Austria}

\author{\sf Ingrid Hotz}
\affiliation{\small\sl DLR Braunschweig, Lilienthalplatz 7, 38108 Braunschweig}

\author{\sf Hans-Christian Hege}
\affiliation{\small\sl Zuse Institute Berlin (ZIB), Takustr.\ 7, 14195 Berlin, Germany}

\author{\sf Bernd R.\ Noack} 
\affiliation{\small\sl
Institut P', 
CNRS -- Universit\'e de Poitiers -- ENSMA, UPR 3346, 
D\'epartment Fluides, Thermique, Combustion,
CEAT, 43 rue de l'A\'erodrome,
F-86036 POITIERS Cedex, France}

\author{\sf Guillaume Daviller}
\affiliation{\small\sl CERFACS, 42 Avenue Gaspard Coriolis, 31057 Toulouse Cedex 01, France}

\author{\sf Marek Morzy\'nski}
\affiliation{\small\sl  Pozna\'n University of Technology, 
Institute of Combustion Engines and Transportation,
 ul.~Piotrowo 3, PL-60-965 Pozna\'n, Poland}

%----abstract-----------------------------------------------------------
\begin{abstract}
In this paper, 
we propose a novel framework to extract features 
such as vortex cores and saddle points in two-dimensional unsteady flows.  
This feature extraction strategy 
generalizes critical points of snapshot topology in a Galilean-invariant manner,
allows to prioritize features according to their strength and longevity,
enables to track the temporal evolution of features,
is robust against noise 
and has no subjective parameters.
These characteristics are realized via several constitutive elements.
First,  acceleration is employed as a feature identifier 
following Goto and Vassilicos (2006),
thus ensuring Galilean invariance.
Second, the acceleration magnitude is used as basis 
for a mathematically well-developed scalar field topology.
The minima of this field are called acceleration feature points,
a superset of the acceleration zeros.
These points are discriminated into vortices and saddle points
depending the spectral properties of the velocity Jacobian.
Third, all operations are based on discrete topology for the scalar field
with combinatorial algorithms.
This parameter-free foundation allows 
(1) to use persistence as a physically meaningful importance measure
to prioritize feature points,
(2) ensures robustness 
since no differentiation and interpolation need to be performed with the data,
and (3) enables a natural and robust tracking algorithm for the temporal feature evolution.
In particular, we can track vortex merging events in an unsupervised manner.
Data based analyses are presented for an incompressible periodic cylinder wake, 
an incompressible planar mixing layer and a weakly compressible planar jet. 
They demonstrate the power of the tracking approach, 
which provides a spatiotemporal hierarchy of the minima.
\end{abstract}

\maketitle 

%----section------------------------------------------------------------
\section{Introduction}
Computational fluid dynamics and particle image velocimetry can provide highly resolved flow data in space and time.
One challenge is to quickly extract the important kinematic features from these data.
Topological methods applied to snapshots are one of the first choices.
Flow topology may provide information about the size of separation bubbles and vortices, about the length of a  dead-water region, and about flow regions, which do not mix --- just to name a few applications.
Velocity snapshot topology provides invaluable insights into laminar or time-averaged flows \cite{Lighthill1963book,Tobak1982arfm,Perry1987arfm,Rodriguez2010jfm,Rodriguez2011tcfd}, or, in general, into velocity fields with a distinguished frame of reference and a low feature density.

Such a topology is always based 
on the zeros of the velocity field
and thus is intrinsically Galilean-variant,
i.e., depends on the chosen frame of reference.
In an unsteady flow, a zero or critical point at one instant is generally not a zero at another instant. 
The question what critical point, 'connector' and other topological elements physically mean for an unsteady situation immediately arises and its answer is far from being clear.

In some cases, e.g., the flow over an obstacle,
a naturally preferred body-fixed frame of reference is given.
Here, Galilean invariance of the topology appears to be a purely academic requirement.
In many cases, however, the proper frame of reference is far less obvious.
In a wake or mixing layer, for instance, 
topology may resolve the initial vortex formation in a body-fixed frame of reference,
but the convecting vortices do not give rise to velocity zeros as they convect downstream.
Now, the choice of the 'right' frame of reference is subject to personal preferences.

A second challenge is that critical points are associated to the smallest structures on the flow. In a fully turbulent flow, the average distance of fixed points is of the order of the Taylor scale~\cite{Wang2006jfm,Wang2008jfm}.
Under these conditions, critical points loose their meaning as 'markers' of large-scale coherent structures.

Third, every measured or simulated data naturally contains a small amount of noise. This noise complicates the extraction of feature points such as zeros. Therefore, important physical structures may be missed. 

To address the first challenge, Goto \& Vassilicos \cite{Goto2006pf} used the acceleration to define a set of feature points. 
They propose to use zeros of the acceleration vector field (\emph{zero acceleration 
points}~(ZAPs)) for the analysis of two-dimensional flows. 
The motivation for the definition of ZAPs was to find a frame moving with vortices, such that the persistence of streamlines is maximized. 
However, also the extraction of physically meaningful zeros of the acceleration is a complex task -- especially in the presence of noise. 

In this paper, we investigate a time-dependent counterpart of the fixed points of 
the velocity field topology. 
Our definition is based on three requirements, namely  
(1) choosing a Lagrangian viewpoint, 
(2) requiring Galilean invariance and 
(3) having standard velocity topology as limiting case for steady flows.
It is shown that the minima of the acceleration magnitude, called \emph{acceleration feature points}, fulfill these criteria.
These points are Galilean-invariant and their physical meaning is inferred from the Jacobian. 
They form a superset of the aforementioned zero acceleration points by Goto \& Vassilicos. 
In contrast to their interesting work, our concept can be generalized to three dimensional flows,
in particular to one-dimensional features. 

The usage of minima enables us to use the powerful concept of scalar field topology and associated combinatorial extraction methods, which are robust against large noise levels in the data.
The application of these methods enables the usage of persistent homology~\cite{Edelsbrunner2008}. 
It serves (a) as a filter for the robust extraction in the first step and (b) as a spatial importance measure for the acceleration feature points. 

A subset of the acceleration feature points can be interpreted as vortex cores. 
Within our combinatorial framework, we track these points over time. 
The combination of persistence with the lifetime of the vortices, we are able to discriminate short-living unimportant features from long-living
and dominant vortices. 
We therefore contribute to the distillation of vortex cores in three major points:
(1) a robust extraction of the feature points in the presence of noise; 
(2) an efficient tracking of them over time;
(3) a filtering strategy that is based on a hierarchy of the vortex cores and trajectories.
The extraction and tracking is based on a combinatorial framework
~\cite{Reininghaus2010a,Reininghaus2011}. The resulting explicit representation of the 
vortex core lines enables a detailed analysis of the interacting structures in a flow field.
In principle, an analogous feature extraction can be effected for saddles.

%With the computation of vortex core lines, we contribute to the extraction of Lagrangian coherent structures (LCS), since vortices are often associated to LCS~\cite{Hussain93}. 
%Another approach that is related to LCS is the extraction of distinguished manifolds
%of particle divergence and convergence. 
%Haller~\cite{Haller2001physd} proposed to extract LCS from ridges of finite-time Lyapunov exponent (FTLE) fields.
%These ridges are also Galilean-invariant and 
%have a clear physical meaning.
%This approach has been applied 
%to several fluid flows \cite{Sadlo2007a,Garth2007,Garth2008b,Chaos2010,Fuchs2010,Tricoche2011}. 
%The price is the introduction of an integration time, a free parameter, which strongly influences the results. 
%In addition, 
%the computational expense for forward and backward integration
%of all flow particles is significant. 
%Our approach does not involve any free parameters, like a time integration window, or a threshold.

This paper is structured as follows: 
In Sec.\ \ref{toc:examples},
key elements of the analysis are motivated 
for simple analytically defined flows.
In Sec.\ \ref{toc:method},
a feature extraction strategy is described.
Results are presented for three planar shear flows
with increasing level of complexity (Sec.\ \ref{toc:results}).
Finally (Sec.\ \ref{toc:conclusions}),
the paper concludes with a summary, the relation to other topological analyses
and an identification of further research questions.
%!TEX root = ./Kasten20111015pf.tex
%----section------------------------------------------------------------
\section{Analytic illustrating examples}
\label{toc:examples}
In this section,
two 2D incompressible flows are considered:
the Stuart solution of the inviscid mixing layer (Sec.\ \ref{toc:Stuart})
and the Oseen vortex pair (Sec.\ \ref{toc:VortexPair}).
These analytical examples show that local minima 
of the total acceleration magnitude
are good indicators of vortices and saddles.
These results motivate the definition 
of acceleration feature points
as key elements of the 
feature extraction strategy elaborated in the next section.
We include a simple three-dimensional flow~(Sec.\ \ref{toc:3Dexample})
in addition to the two-dimensional examples.

%----subsection---------------------------------------------------------
\subsection{Stuart solution of the mixing layer}
\label{toc:Stuart}
An incompressible mixing layer is described
in a Cartesian coordinate system 
$\mathbf{x}=(x,y)$,
where  $x$  and $y$ represent 
the streamwise and transverse coordinate, respectively.
The origin $\mathbf{0}$ is placed in one saddle.
The velocity is denoted by $\mathbf{u}=(u,v)$,
where $u$ and $v$ represent its $x$ and $y$ components, respectively.
All quantities are normalized 
with half of the relative velocity difference
and half of the vorticity thickness. 

Targeting a simple analytical example,
we consider a streamwise periodic mixing layer
with constant width, 
as described by the inviscid Stuart solution  \cite{Stuart1967jfm}:
\begin{subequations}
\label{eqn:Stuart}
\begin{eqnarray}
u &=& u_c+\frac{\sinh(y)}{\cosh(y)\!-\!0.25\cos(x\!-\!u_c t)}, \\
v &=& \!-0.25\cdot\frac{\sin(x-t)}{\cosh(y)\!-\!0.25\cos(x\!-\!u_c t)},
\end{eqnarray}
\end{subequations}
where $u_c$ represents the convection velocity.

%----figure--------------------------------------------------------------
\begin{figure}
%\vspace{10mm}
\begin{center}
\includegraphics[width=0.49\textwidth]{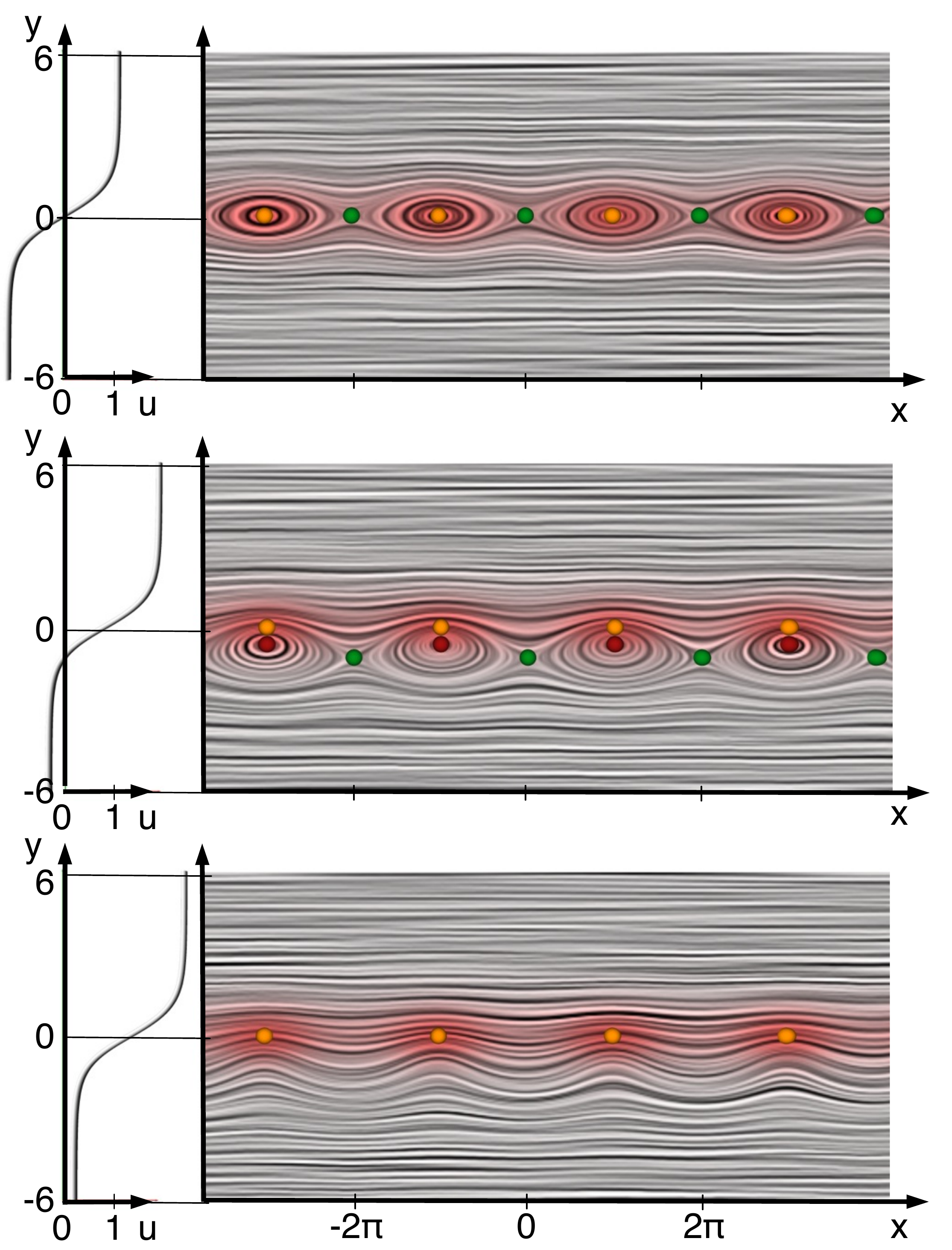}
\end{center}
\caption[X]{Stuart vortices in various convecting frames. The mean velocity profile is shown 
at the left. The Stuart vortices are depicted by visualizing the instantaneous velocity field 
using line integral convolution.
The coloring is determined by vorticity; more intense red corresponds to higher vorticity.
The critical points of standard velocity field topology  are displayed as red (centers) and green (saddles) spheres.
The maxima of the vorticity are added as orange spheres.
}
\label{fig:Stuart}
\end{figure}
%-----------------------------------------------------------------------
The  Stuart vortices are depicted in Fig.\ \ref{fig:Stuart}
as streamlines using planar line integral convolution (LIC) \cite{Cabral:1993,Stalling:1995}.
The top picture represents Eqs.\ \eqref{eqn:Stuart}
and shows the famous cat eyes
in a periodic sequence of centers (vortices) and saddles
for a vortex-fixed frame of reference ($u_c=0$).
The middle picture depicts the same structures
but in a frame of reference moving to the left 
with the lower stream at velocity $(-0.7,0)$, or, equivalently,
the vortices moving to the right at $u_c=0.7$.
The centers and saddles are displaced towards the slower stream.
The bottom picture illustrates the same flow
with a frame of reference moving 
at velocity $(-1.2,0)$,
i.e.\ $u_c=1.2$ in Eq.\ \eqref{eqn:Stuart}.
Now, no zeros are observed.
These pictures recall the well-known fact, emphasized in many textbooks in fluid mechanics,
that velocity field topology strongly depends on the frame of reference, i.e.\ is not Galilean-invariant.
% This is emphasized in many text books in fluid mechanics, also for other flows.
In case of the Stuart solution,
one might argue that the frame of reference convecting
with the structures is the most natural one.
However, the convection velocity of a jet
and many other flows depend on the streamwise position,
i.e., generally no single natural frame of reference exists
for topological considerations.

\begin{figure}[!htb]
%\vspace{10mm}
\begin{center}
\includegraphics[width=0.49\textwidth]{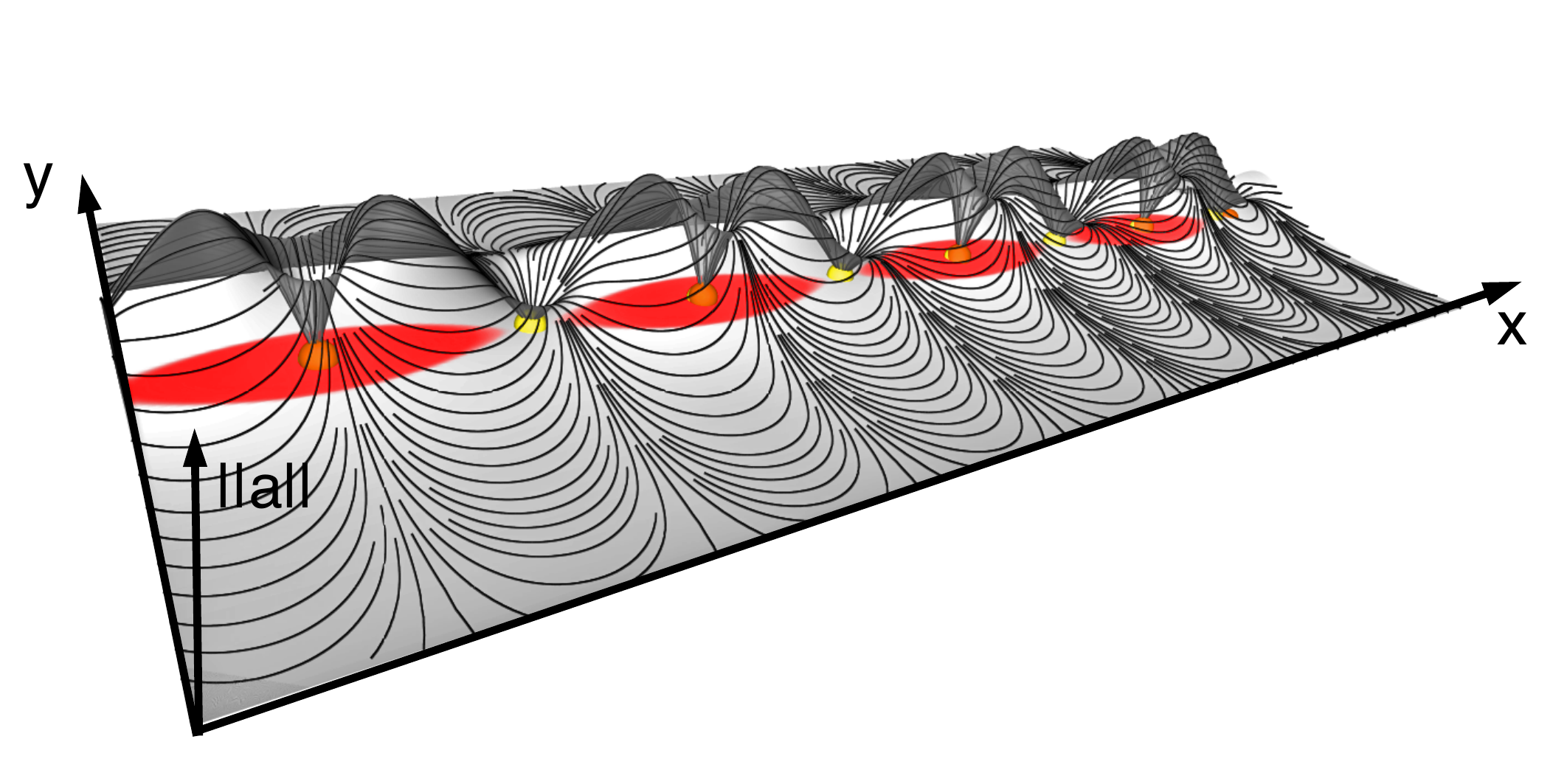}
\end{center}
\caption[X]{The acceleration field of convecting Stuart vortices. The coloring at the 
bottom is determined by the vorticity. The height field shows the acceleration magnitude
and the curves depict integral lines of the acceleration vector field. The yellow spheres
highlight the acceleration minima, orange spheres vorticity maxima. Note that 
center-like acceleration minima and vorticity maxima coincide (orange spheres hide yellow spheres).
}
\label{fig:stuart-acceleration}
\end{figure}

The saddles and centers of a Stuart solution 
are not only zeros of the velocity field
but also zeros of the material acceleration field
\begin{equation}
\mathbf{a}=D_t \mathbf{u}= \partial_t \mathbf{u} + \mathbf{u} \cdot \nabla \mathbf{u}.\label{acceleration-of-particle}
\end{equation}
Here, $\partial_t$ represents the partial derivative with respect to time,
$\nabla$ the nabla operator and the dot $\cdot$ the tensor contraction.
The acceleration zeros are derived from a Galilean-invariant field
and do not depend on the chosen inertial frame of reference.
Figure~\ref{fig:stuart-acceleration} illustrates the acceleration field as height field.
%----figure--------------------------------------------------------------
%-----------------------------------------------------------------------
The zeros of the acceleration field  and
the local minima of the acceleration magnitude (yellow spheres) coincide
in this example.
In general, 
the latter quantity is a superset of the first.
The acceleration minima, however, 
enable to identify vortices and saddles
in case of a non-uniform convection velocity.

\subsection{Oseen vortex pair}
\label{toc:VortexPair}
In this section, 
a pair of equal Oseen vortices in ambient flow is considered.
They rotate around the origin 
at constant distance $R=1 / \sqrt{2}$
with uniform angular velocity $\Omega$.
Let $\mathbf{x}_i = (x_i,y_i)$, $i=1,2$ be the centers
of the two vortices.
Then, 
\begin{subequations}
\label{eqn:VortexPair}
\begin{eqnarray}
x_1  =&  \!R \> \cos \Omega t, \quad &y_1 = R \> \sin \Omega t, \\
x_2  = &\!-R \> \cos \Omega t, \quad &y_2 = - R \> \sin \Omega t. 
\end{eqnarray}
\end{subequations} 
The induced velocity $u_{\theta}$ of a single Oseen vortex 
in the circumferential direction $\theta$ is given by 
\begin{equation}
%\begin{eqnarray}
 u_{\theta}(r) = \frac{\Gamma}{2\pi\> r} \> \left( 1-e^{ - \left( r / r_c \right)^2}  \right), 
%\end{eqnarray}
\label{eqn:Vortex}
\end{equation}
where $r$ is the distance from the center of the vortex, 
$r_c$  determines the core radius and
$\Gamma$ is the circulation of the vortex. 
%----figure--------------------------------------------------------------
\begin{figure}[!htb]
%\vspace{10mm}
\begin{center}
\includegraphics[width=0.49\textwidth]{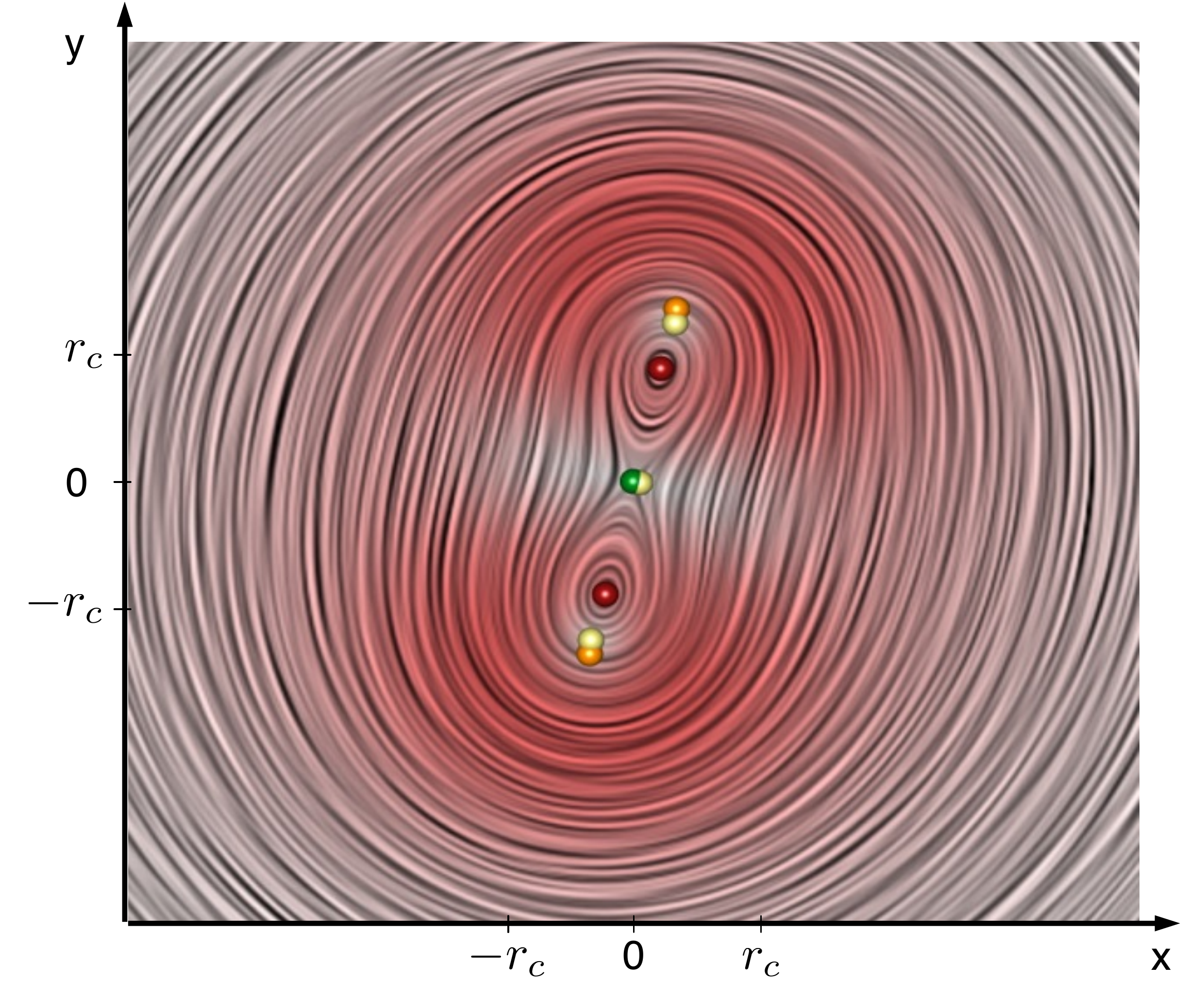}
\end{center}
\caption[X]{Depiction of two co-rotating Oseen vortices by their instantaneous streamlines using line integral convolution.
The color-coding is determined by the acceleration magnitude; more intense red encodes higher values.
The minima of the acceleration magnitude are marked by yellow spheres.
For comparison, the critical points of standard velocity 
field topology (green and red spheres) and the maxima of the vorticity (orange spheres) are added.}
\label{fig:VortexPair}
\end{figure}
%-----------------------------------------------------------------------
In our example, 
$r_c$ is chosen as $0.5$ and $\Gamma$ as $2\pi$.
The velocity of each vortex, cf. Eq.\ \eqref{eqn:VortexPair},
corresponds to the induced velocity, cf. Eq.\ \eqref{eqn:Vortex},
i.e.\ $R \Omega = u_{\theta}(2R)$.
The diffusion of vorticity is ignored 
to keep this analytical example simple.
Figure~\ref{fig:VortexPair} illustrates 
the Oseen vortex pair as described by Eq.\ \eqref{eqn:VortexPair} and Eq.\ \eqref{eqn:Vortex}.
Here, the instantaneous streamlines can be inferred from the planar LIC.

Red and green spheres mark the corresponding critical points of the velocity field.
Regions of large acceleration are marked by red.
The minima (zeros) of the acceleration field are denoted by yellow circles.
It should be noted that 
the maxima of the vorticity (indicated by orange spheres)
are much closer to the minima (zeros) of the acceleration
than to the zeros of the velocities.
This difference is insignificant for 'frozen' vortices
but increases with increasing angular rotation.
Hence, the difference is correlated 
with the radial acceleration of the vortex motion.
Only a non-inertial co-rotating frame of reference 
minimizes the difference between vorticity maximum
and the center of the velocity field.

It may be noted that any definition of feature points is based on the concept of an 'idealized template',
like a slowly accelerating saddle or vortex 
--- as compared to the surrounding acceleration maximum. 
And any definition can be challenged by constructed limiting examples.
In the example of an Oseen vortex pair,
the limit of infinitely fast rotation of the vortices
around their center constitutes such an example.
However, this kinematic example would violate the Biot-Savart law,
i.e.\ is not physically realizable.

\subsection{Simple 3D flow}
\label{toc:3Dexample}
%----figure--------------------------------------------------------------
\begin{figure}[!htb]
%\vspace{10mm}
\begin{center}
\includegraphics[width=0.49\textwidth]{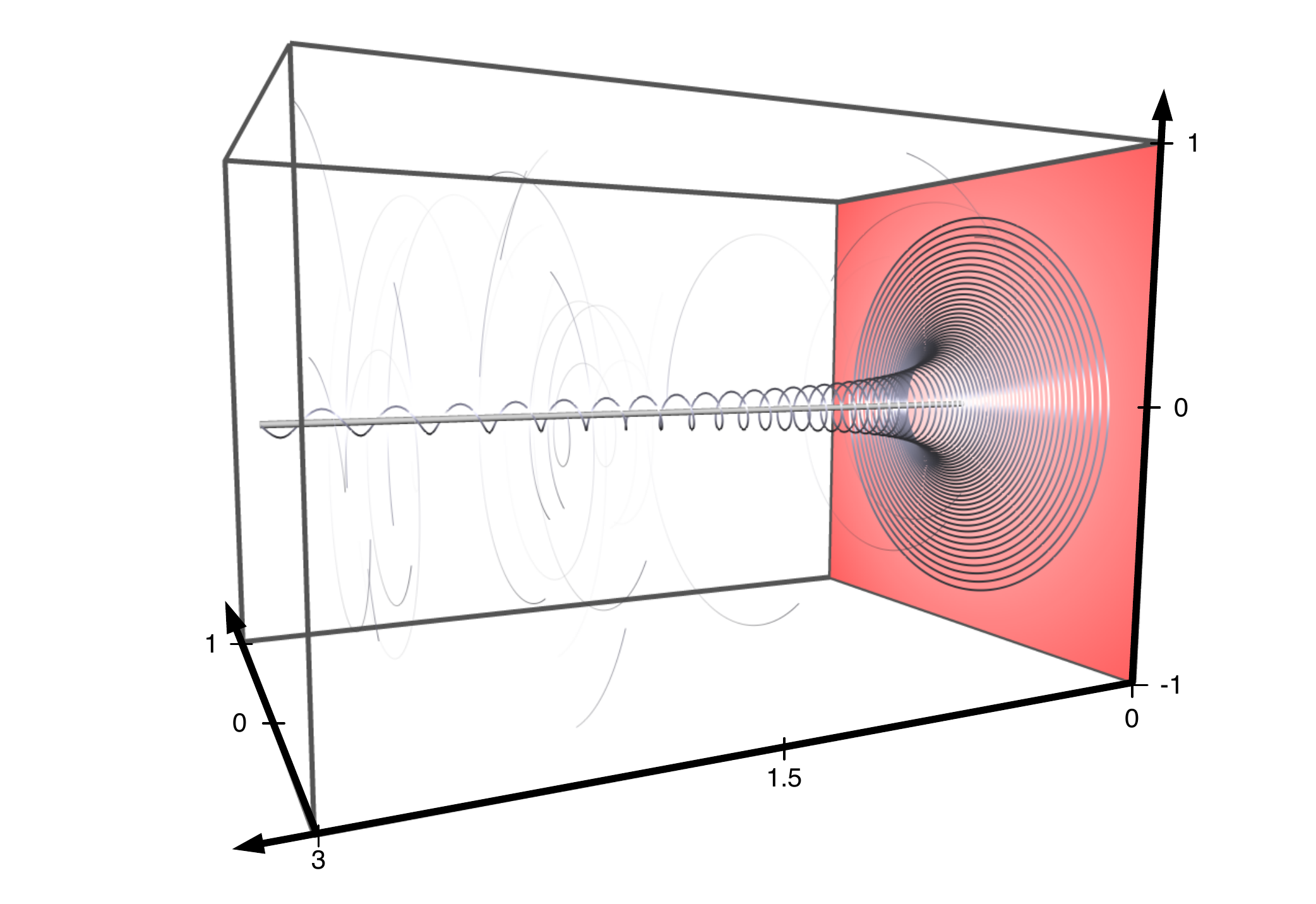}
\end{center}
\caption[X]{Depiction of a simple three-dimensional stationary flow field. The 
particle trajectories are indicated by the illuminated streamlines. 
In the center, the vortex core line is extracted as minimal line of the acceleration magnitude. 
Note that the particles along the vortex core line have non-zero acceleration.}
\label{fig:3DExample}
\end{figure}
%-----------------------------------------------------------------------
%Finally, we consider a simple three-dimensional flow field. 
Following Wu et al.~\cite{Wu2005axial}, we consider a simple linear velocity field with swirl and strain.
The velocity field is given by 
\begin{eqnarray}
u &=& -0.75 x-100 y, \nonumber\\
v &=& -0.75 y+100 x,  \label{eqn:3DExample}\\
w &=& 1.5 z.\nonumber
\end{eqnarray}
The material acceleration $\mathbf a=(a_u,a_v,a_w)$ of this field is given by
\begin{eqnarray}
a_u &=& -9999.4375 x+150 y, \nonumber\\
a_v &=& -9999.4375 y-150 x,\label{eqn:3DExampleAcceleration}\\
a_w &=& 2.25 z.\nonumber
\end{eqnarray}
In Fig.~\ref{fig:3DExample}, some streamlines of this field are randomly placed in the volume.
One streamline is emphasized showing swirling motion around the vortex core line (z-axis). 
%The acceleration magnitude is shown as  color map on the back face of the volume as reference.
The points on the core line are minima of the acceleration magnitude of the flow field with respect to 
a two-dimensional cross-section. 
Using  the terminology of  three-dimensional scalar field topology, such lines
are called \emph{valley lines} or \emph{minimal lines}.  They can be extracted from the acceleration magnitude field 
without prior knowledge of the corresponding cross-sections.
In this example, the values of the acceleration are non-zero everywhere  along the line.
While for two-dimensional fields, the centripetal acceleration of a vortex always induces a zero point, 
this characterization is not transferable to the three-dimensional case.
Now, minimal lines provide a meaningful generalization of the two-dimensional feature points to the 
three-dimensional setting.

%!TEX root = ./Kasten20111015pf.tex
%----section------------------------------------------------------------
\section{Feature Extraction}
\label{toc:method}
\begin{figure}
%\vspace{10mm}
\begin{center}
\includegraphics[width=0.35\textwidth]{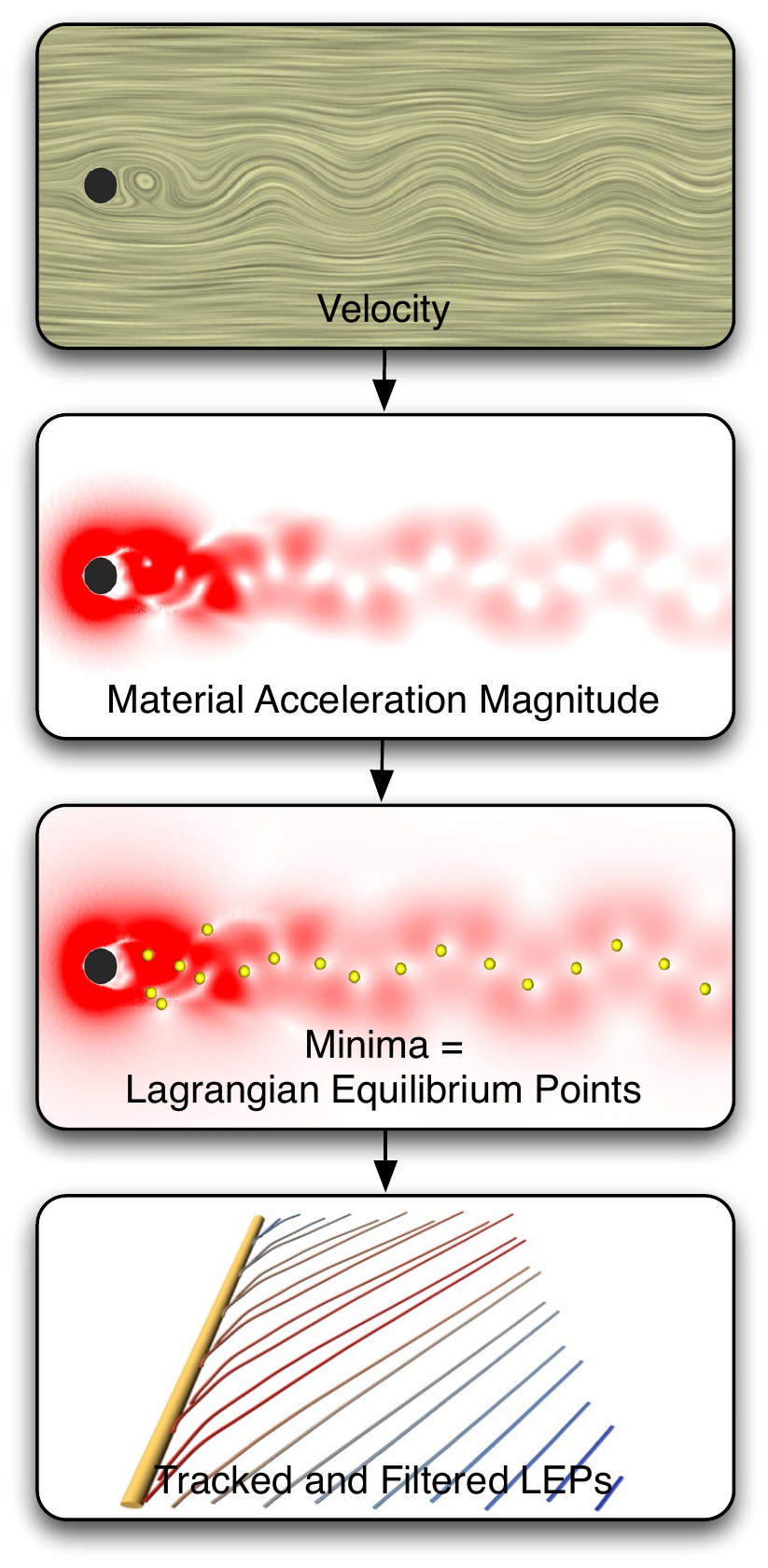}
\end{center}
\caption[X]{Pipeline of the proposed approach: After computing the acceleration
magnitude field from the velocity, 
its minima are extracted, 
which are referred to as \emph{Lagrangian Equilibrium Points} (LEPs).
These LEPs are tracked over time and prioritized 
by a spatiotemporal importance measure. 
The importance measure combines a spatial strength and the lifetime 
of the feature.}
\label{fig:Pipeline}
\end{figure}

Our feature extraction pipeline consists of three steps: 
(1) spatial feature extraction, resulting in isolated feature points;
(2) computation of the temporal evolution of these points; and
(3) spatiotemporal filtering of the resulting structures, cf.\ Fig.\ref{fig:Pipeline}.

First (Sec.~\ref{subsec:LEP}), distinguished feature points are defined.
One important implication of this definition for the pressure field 
is described in  Sec.~\ref{subsec:implications}.  
Finally, we describe a mathematical model
for the extraction and tracking algorithm   (Sec.~\ref{subsec:strategy}).

%----subsection---------------------------------------------------------
\subsection{Acceleration feature points}
\label{subsec:LEP}
In this section, the definition of the considered feature is introduced.
Starting point is a critical review of the velocity snapshot topology. 
Topological analysis of velocity fields has been successfully applied for 
examination of flow fields with a distinguished frame of reference.
However, its applicability is limited, as location and number of critical points depend on the frame of reference.
The goal of the current study is a definition of an alternative feature concept,
which generalizes the snapshot topology in a local sense 
and overcomes the above-mentioned limitations.
The feature point definition is motivated by the observations in the previous examples and the following three requirements:

\begin{description}
\item{\emph{(R1)} Correspondence to velocity topology:}
A flow field is called steady, if there exists a distinguished frame of reference 
for which the vector field is stationary, i.e., it does not change in time. 
Such flow fields consist of \emph{frozen} convective structures. 
They satisfy Taylor's hypothesis~\cite{taylor1938spectrum}. 
With respect to this distinguished frame of reference, critical points of the velocity field correspond to the position of vortex cores and saddles.
This concept is not applicable to unsteady flow fields, since there is no such distinguished 
frame of reference. 
Aiming for a generalization of velocity topology, the newly defined feature points should coincide for steady flow fields with the zeros of the velocity field. This also means that  the classification 
of the points as saddles or centers is preserved. 
Note that this requirement is not fulfilled by Haller's definition of an objective vortex~\cite{Haller2005jfm}. 
Rotational invariant features cannot distinguish saddles and centers. 
%We consider a class of flows
%with 'frozen' convective structures,
%i.e.\ velocity fields satisfying Taylor's hypothesis \cite{taylor1938spectrum}
%with a convection velocity $u_c$ in $x$-direction:
%$$ \mathbf{u}(\mathbf{x},t) 
%   = u_c \> \mathbf{e}_x + \mathbf{f}(\mathbf{x}- u_c  t \> \mathbf{e}_x ).$$
%Here, $\mathbf{e}_x$ is a unit vector pointing in positive $x$-direction. 
%We require that the new feature point $\mathbf{x}_0$ 
%is identical to the corresponding zero of the convected system,
%$$ \mathbf{u}^{\prime}(\mathbf{x}_0^{\prime})=\mathbf{f} \left ( \mathbf{x}_0^{\prime} \right) = 0,$$ 
%after the coordinate transformation
%\begin{eqnarray*}
%\mathbf{x}'&=& \mathbf{x}-u_c\, t \> \mathbf{e}_x ,\\
%\mathbf{u}'&=& \mathbf{u}-u_c\,   \> \mathbf{e}_x .
%\end{eqnarray*}
%Here, $u_c$ is the convection velocity in direction $\mathbf{e}_x$.

\item{\emph{(R2)} Galilean invariance:}
A Galilean-invariant feature identifier reveals the same structures when changing the frame of reference.
\begin{figure}%
    \centering
        \subfigure[Convecting center]{\includegraphics[width=0.49\linewidth]{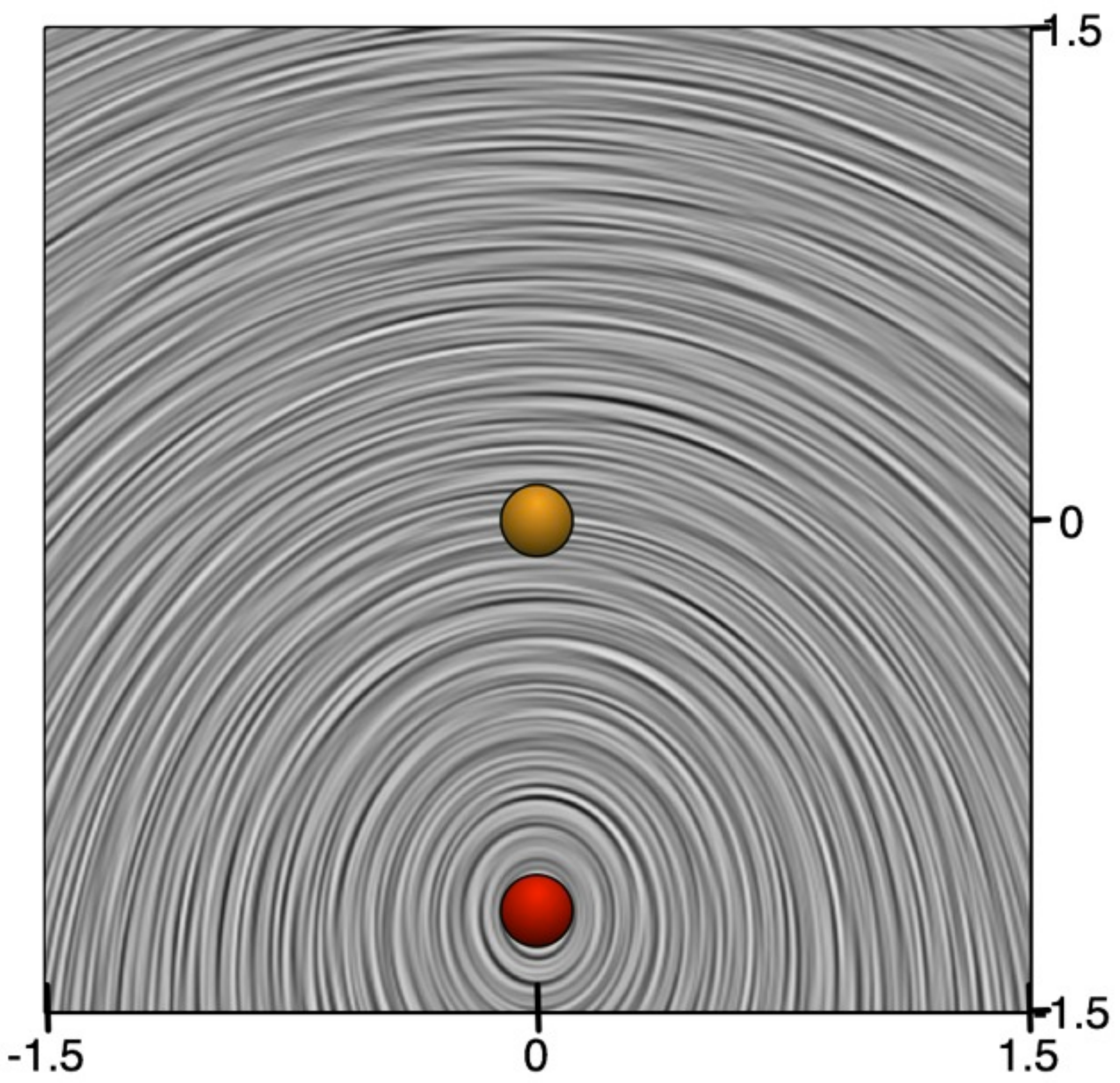}}
        \subfigure[Distinguished frame of reference]{\includegraphics[width=0.49\linewidth]{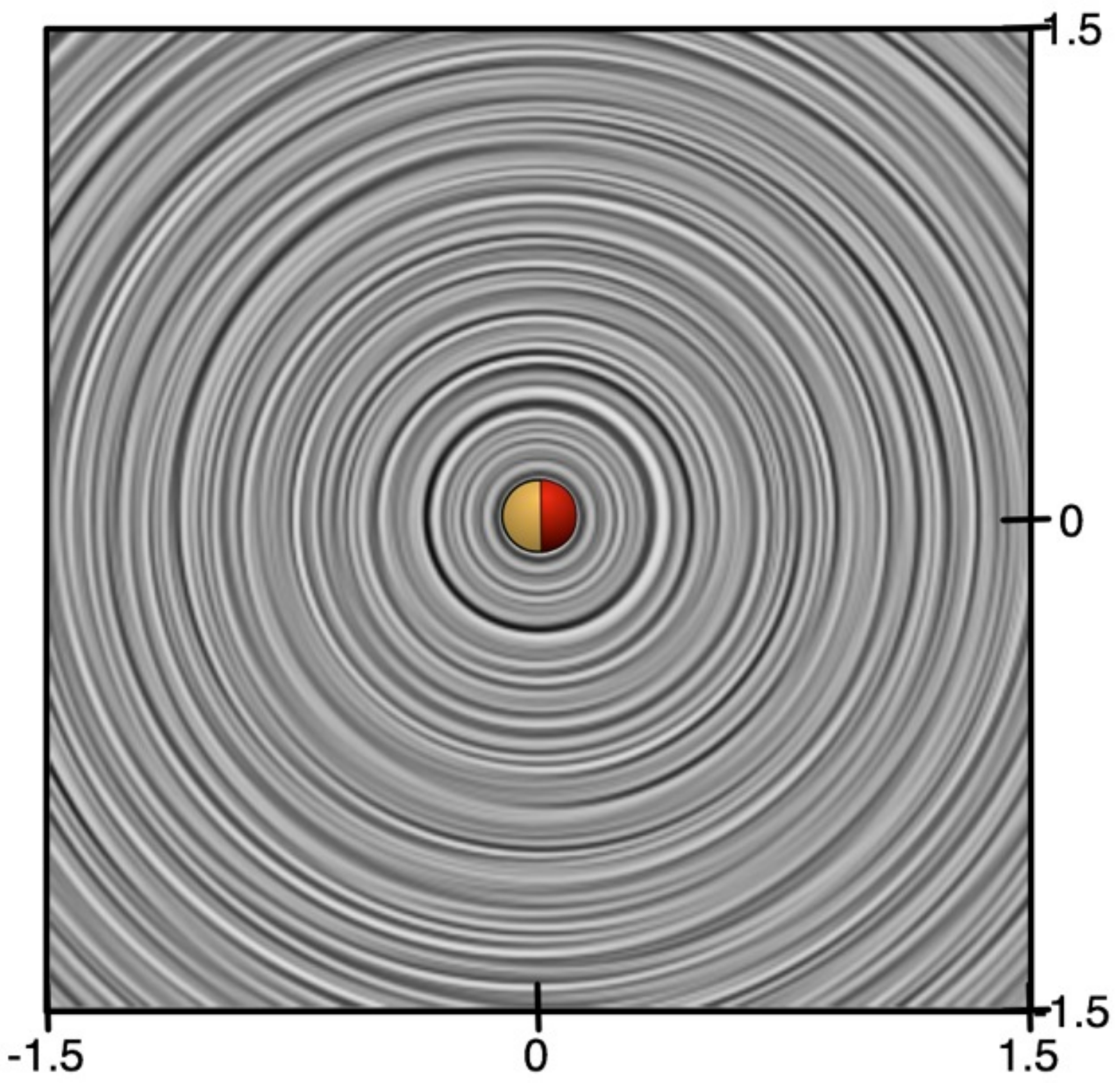}}
    \caption[Frames of reference]{Convecting Oseen vortex displayed with respect to two
    different reference frames. A dense streamline visualization builds the background texture for both images.
    The maxima of the vorticity are marked by orange spheres. They coincide with the acceleration minima which are therefore not       
    shown in both images.  
    The fixed-points of the velocity field are represented as a red sphere.
    The vorticity maximum reveals a vortex core moving along the x-axis. 
    In image (a), this point does not coincide with the center of  the streamlines.
     Applying a Galilean transformation such that the resulting flow is steady leads to a convected frame of reference~(b)
     revealing a different picture. Now the fixed point coincides with the maximum of vorticity and the minimum of the 
     acceleration.}\label{fig:LEP_def_convCenter}
\end{figure}%
An example illustrating the influence of a Galilean transformation on the velocity field and its feature points is illustrated 
in Fig.~\ref{fig:LEP_def_convCenter}.
%we will have a closer look on the meaning of a Galilean transformation for the purpose of 
%getting a better understanding of Galilean invariance and frames of reference. 
Let $\mathbf v(\mathbf x,t)$ be a time-dependent vector field %on a finite spatial domain $D\subset \mathbb{R}^2$ 
with a single Oseen vortex, Eq.~\ref{eqn:Vortex} with $r_c=2$, $\Gamma=8\pi$, convecting with constant velocity 
$\left(x_0(t)=v_0 t, y_0(t)=0\right)$ from left to right. The instantaneous velocity field for one point in time is shown in Fig.~\ref{fig:LEP_def_convCenter}~(a) including the maximum of the vorticity~(orange) and the fixed point~(red). The two points do not coincide. Since this is a steady flow field, there exists a distinguished frame of reference moving with the center, that can be reached by a Galilean transformation.
The resulting velocity field is shown in Fig.~\ref{fig:LEP_def_convCenter}~(b). 
While the flow behavior itself is not changed by the transformation, the visual output is different. In particular, both feature points now coincide.

\item{\emph{(R3) Lagrangian viewpoint}:}
%Classification of feature points:}}
To guarantee a physically sensible feature identifier, 
we focus on particle motion. 
This Lagrangian viewpoint implies the focus on Galilean-invariant properties of fluid particles,
but it does not include tracking finite-time fluid particle motion.
This restricted Lagrangian viewpoint is consistent 
with the general notion of 'Lagrangian coherent structures.'
%BRN120711: Note that I added some sentences.
%The classification of non-degenerate critical points  must be preserved for incompressible velocity field topology (saddles and centers). 
\end{description}

These requirements and the observations from Sec.~\ref{toc:examples} 
suggest to relate feature points to the material acceleration field. 
The particle acceleration $\mathbf{a}$ is the total derivative of the
flow field ${\bf u}$. 
In other words, the  acceleration in a space-time point
$({\mathbf x}, t)$ is given by Eq.~\eqref{acceleration-of-particle}.

\emph{Definition:}  A minimum of the magnitude of the material  acceleration $\|{\mathbf a}\|$ is called 
\emph{acceleration feature point}. 
Such points can be classified on the basis
of the Jacobian of the velocity field, $\nabla \mathbf{u}$.  
A feature point is called \emph{saddle-like} if its eigenvalues are real and 
\emph{center-like} if its eigenvalues are complex.
A feature trajectory is defined by the temporal evolution of a minimum in the acceleration field.

In the following, this definition is shown to satisfy requirements R1 to R3.
Let $\mathbf{x}_0$ be a zero of the velocity field $\mathbf{u}(\mathbf{x}_0,t) \equiv 0$.
This implies that the material acceleration 
$\mathbf{a} \vert_{\mathbf{x}_0} = 
\left( \partial_t \mathbf{u} 
     + \mathbf{u} \cdot \nabla \mathbf{u} 
\right)  \vert_{\mathbf{x}_0}= 0$
vanishes at $\mathbf{x}_0$.
Thus, the minima of the acceleration magnitude are a superset of acceleration zeros
and these zeros are a superset of the critical points of the velocity field.
Hence, acceleration feature points can be considered to be a generalization of the critical points of the velocity fields. 
Acceleration is a Galilean-invariant quantity. It is computed from the velocity using the material or
Lagrangian derivative that links the Eulerian to the Lagrangian viewpoint~\cite{Panton1984book}.
Moreover, acceleration feature points satisfy all requirements R1 to R3.

The acceleration feature points can exhibit vortex- as well as saddle-like behavior,
depending on the eigenvalues of the velocity Jacobian. 
Real eigenvalues correspond to saddles while a complex-conjugate eigenvalue pair
indicates a vortex.
Alternative synonymous discriminants have been proposed for 2D flows:
Goto \& Vassilicos \cite{Goto2006pf} show that saddles are associated
with sources of the material acceleration field
while vortices correspond to sinks.
One advantage of their definition is that it relies purely on the acceleration
without reference to the velocity Jacobian.
Basdevant \& Philipovitch \cite{Basdevan1994pd}
critically assess the use of the Weiss criterion
as discriminant.

\subsection{Implications of acceleration feature points}
\label{subsec:implications} 
Another perspective onto the acceleration minima is given by their relation to
the pressure gradient via the incompressible Navier-Stokes equation:
\begin{equation}
	\begin{aligned}
 {\mathbf a}({\mathbf x}, t) 	& = - \frac{1}{\rho} \> \nabla p({\mathbf x},t) 
	+ \nu \, \Delta {\mathbf u}({\mathbf x},t), \label{eqn:ns-1}\\
         					    0 	& = \hspace{7pt} \nabla \cdot {\mathbf u}({\mathbf x},t) \text{,}
	\end{aligned}
\end{equation}
where $p$ is the pressure of the flow field, 
$\rho$ and $\nu$ are the kinematic viscosity and density of the fluid, respectively, 
and $\Delta$ is the spatial Laplacian operator. 
For ideal flows, the equations reduce to the Euler equation: 
\begin{equation}
	\begin{aligned}
		{\mathbf a}({\mathbf x}, t) 	& =  -\frac{1}{\rho} \> \nabla p({\mathbf x},t) ,\\
                          		0	 		& =   \hspace{7pt}  \nabla \cdot {\mathbf u}({\mathbf x},t) \text{.}
	\end{aligned}
\end{equation}
Then, local extrema of the pressure field, which are zero points of the pressure gradient
coincide with zeros of the acceleration field. 
In this case, the above defined acceleration feature points form a superset
of local extrema of the pressure field, 
the minima of which are often associated with vortices.
%It should be noted the same pressure Poisson equation
%can be derived for the Navier-Stokes and Euler equation,
%indicating the 

\subsection{Feature Point Extraction Strategy}
\label{subsec:strategy} 
A second task, besides the definition of physically appropriate feature points,
is the selection of a mathematical model that serves as basis for an algorithmic implementation.
The selection of a mathematical framework is guided by the following criteria:
\begin{description}
\item{\emph{(C1)}} It should facilitate a robust and efficient extraction without subjective parameters
to enable an unsupervised extraction of the structures. 
 \item{\emph{(C2)}} It should allow to generate a feature hierarchy based on an intrinsic filtering mechanism. This eases the interpretation of the results and becomes necessary as soon as one moves on from simple showcases or when the data exhibit high feature densities.
 \item{\emph{(C3)}} It should allow for tracking of features over time, based on neighborhood relations.
\end{description}
From a mathematical point of view, extrema and ridges of scalar fields can be subsumed 
under the framework of scalar field topology.
This provides access to many powerful algorithmic tools developed for extraction and tracking of topological  structures in scalar fields.
The algorithm to extract minima of scalar fields used in this paper is based 
on \emph{discrete Morse theory}~\cite{Forman1998b}.
It is purely combinatorial and guarantees topological consistency of the extracted 
structures~\cite{Reininghaus2010a}.
The robustness of the algorithm and lack of any algorithmic parameter allow an unsupervised extraction of structures.
This guarantees the applicability of the methods to large data sets. 
In the following, we will briefly describe the used methods and summarize the concept of discrete Morse theory. 
\\
\\
\emph{Combinatorial extraction of two-dimensional scalar field topology} -- 
Typically, scalar field topology is extracted by analyzing the gradient of the scalar function. This approach utilizes derivatives that amplify noise when computed in a discrete setting. In contrast, combinatorial methods do not rely on interpolation or derivatives. Therefore, we decided to use a combinatorial  setting, here, following the ideas of Reininghaus et al.~\cite{Reininghaus2010a}. Note that our data is given on a polygonal grid for each time slice. 

In the following, we briefly describe how to extract scalar field topology using the aforementioned approach: The grid that holds the data is transferred to a polygonal graph. In this graph, each node represents a $d$-dimensional face (point, edge, polygon) and each link represents the connection between two adjacent faces. In accordance to Forman~\cite{Forman1998b}, the combinatorial gradient field is given as a matching on this graph. The critical points are represented by the unmatched nodes and the streamlines are alternating paths with respect to the matching. An optimization algorithm enforces the matching to represent the actual analytic gradient field best. Loosely speaking, it has to be assured that a combinatorial stream line corresponds best to the analytical stream lines. A couple of algorithmic implementations of this idea have been proposed; we follow the approach presented by Robins et al.~\cite{robins10}.
\\
\\
\emph{Generating a feature hierarchy for acceleration feature points} -- 
One way to approach the problem of a high feature density are statistical methods. 
Another way, pursued in this work, is to facilitate an importance measure to build a feature hierarchy.
A commonly used importance measure for critical points in context of scalar field topology is
\emph{persistent homology}~\cite{Edelsbrunner2008}. 

Persistence measures the stability of critical points with respect to changes in the data as introduced by noise, for instance.
To do so, persistence analyzes the topological changes of the sublevel sets of a scalar function. 
At critical points, the topology of the sublevel sets changes by increasing the sublevel parameter. 
In two dimensions, there are four events possible: First, a new connected component in the sub-level set can be born. This happens at a minimum. Furthermore, two connected components can merge, which occurs when the parameter $t$ passes the value of a saddle. Third, a new hole can also be born at a saddle. Last, a hole in a connected component can die. This occurs at a maximum. 

The persistence value of the critical points is determined in the following way: A new-born connected component is labeled with the associated minimum. At a saddle, two connected components merge that are labeled with two different minima. The persistence value for the saddle and the minimum with the higher scalar value is defined as their scalar value difference. The merged component is labeled with the remaining minimum. A maximum is paired with the saddle with the highest function value and, again, their persistence value is defined as their scalar value difference.

Algorithmically, persistence can be computed in two dimensions by an approach that can also be used to compute the matching, see Ref.~\cite{Reininghaus2010a}.  For two-dimensional fields of reasonable resolution, the implementation computes persistence in a few seconds. This enables us to analyze time-dependent fields with many time slices. The advantage of this algorithm is that it not only computes the persistence values for the critical points but it is also able to simplify the combinatorial gradient field following the hierarchy introduced by persistence. This can be used to reduce noise and therefore facilitates the tracking of the critical points in the next step. 

\begin{figure*}
%\vspace{10mm}
\begin{center}
\subfigure[]{
	\includegraphics[height=0.7\columnwidth]{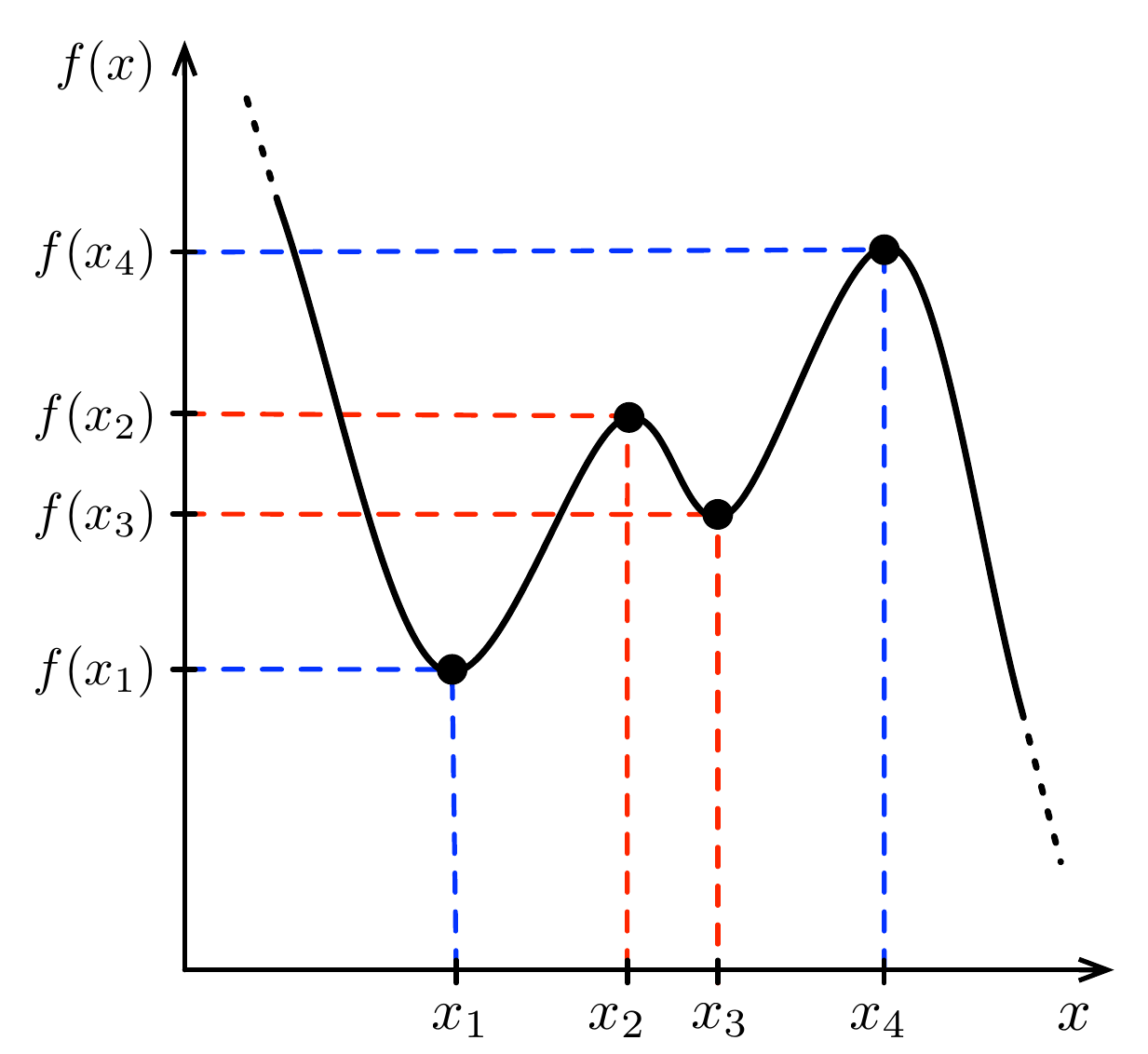}\label{fig:persistenceexplain}
}
\subfigure[]{
	\includegraphics[height=0.7\columnwidth]{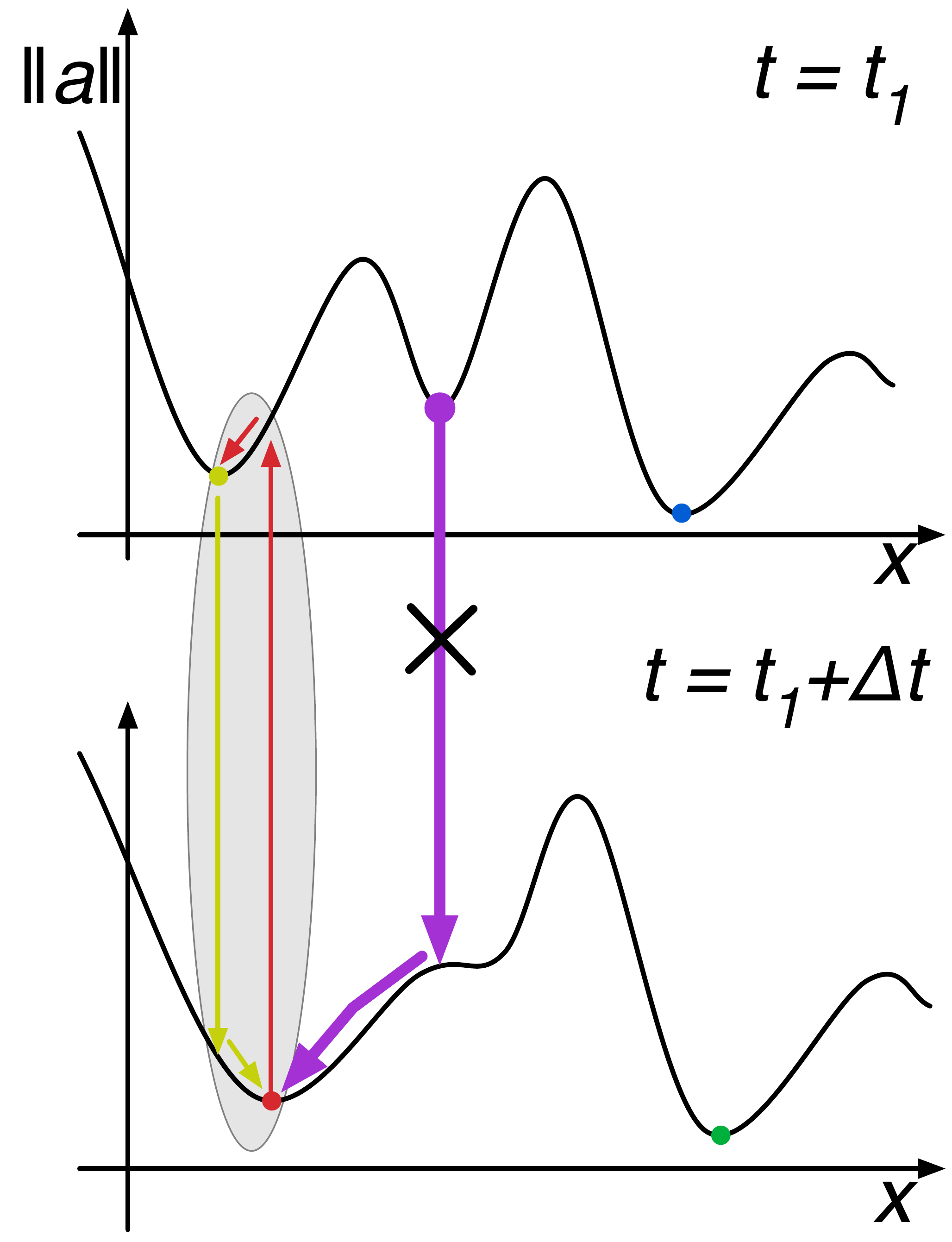}\label{fig:tracking}
}
\caption{
(a): The employed spatial feature importance is given by the persistent homology of the critical points. It measures how strong a minimum is, compared 
to its neighbors. This is achieved by correctly pairing critical points and
quantifying their height difference. The image shows such a pairing, e.g., the critical 
points at $x_2$ and $x_3$. (b): To extract the evolution of the acceleration feature points in time, the minima are tracked. The applied method utilizes the concept of combinatorial feature flow fields (CFFF). In this setting, two minima of adjacent time slices are uniquely connected, 
if they lie in the basin of each other.}
\end{center}
\end{figure*}

Note that critical points that are paired by persistence are not necessarily adjacent. For a simple example illustrating the concept of persistent homology, we refer the reader to Figure~\ref{fig:persistenceexplain}.
\\
\\
\emph{Temporal feature development} -- 
To get an understanding of the temporal evolution of acceleration feature points, the minima are tracked over time.
In the context of discrete Morse theory, one can make
use of {\emph combinatorial feature flow fields} (CFFF), as proposed in Ref.~\cite{Reininghaus2011}. 
The basic idea of this tracking algorithm is to construct a
discrete gradient field in space-time, describing the development
of the acceleration minima, such that tracking of those minima results in
an integration of the discrete gradient field. 
The approach constructs two fields: A forward and a backward field representing the destination of a critical point in forward or backward direction, respectively. Two minima of two adjacent time slices are \emph{uniquely} connected to each other, if they are connected in the forward as well as in the backward field. Intuitively, these two  minima are uniquely connected if they fall into each other's basin, see Figure~\ref{fig:tracking}.
The result of this tracking is a set of temporal feature lines without mergers and splits.
By further incorporating the aforementioned spatiotemporal gradient fields, we were able to extract the mergers that occur for vortex core lines in a two-dimensional setting. To do so, we utilize the forward tracking field to connect the end of a uniquely tracked minima line to another minima line in space time. The general approach is described in~\cite{Kasten2012b}.
\\ \\
%
%----figure--------------------------------------------------------------
%-----------------------------------------------------------------------
%
\emph{Generating a feature hierarchy for temporal feature lines} -- 
A typical importance measure applied for time dependent features 
is the concept of \emph{feature lifetime}.
While leading to much simpler results, such filter methods are purely based 
on temporal measures. They ignore pronounced short-lived features, 
which can play a significant role for the flow.
Therefore, the temporal importance measure, given by the feature lifetime, should be combined with a spatial feature strength.
The CFFF approach allows for a straightforward incorporation of persistent homology
as spatial importance measure for feature lines.
In this paper, the spatiotemporal importance measure is defined by  
integrating persistence along the feature line, e.g., by accumulating
all persistence values along the line. Since the measure is
not normalized, the lifetime of the
feature is inherently depicted by this measure -- longer living features
are more important if all structures have the same spatial strength.
Using this importance measure, it is possible to filter out short-living weak
features.

%!TEX root = ./Kasten20111015pf.tex
%----section------------------------------------------------------------
\section{Results for free shear flows}
\label{toc:results}
%----figure--------------------------------------------------------------
\begin{figure}[htb!]
%\vspace{10mm}
\begin{center}
\includegraphics[width=0.49\textwidth]{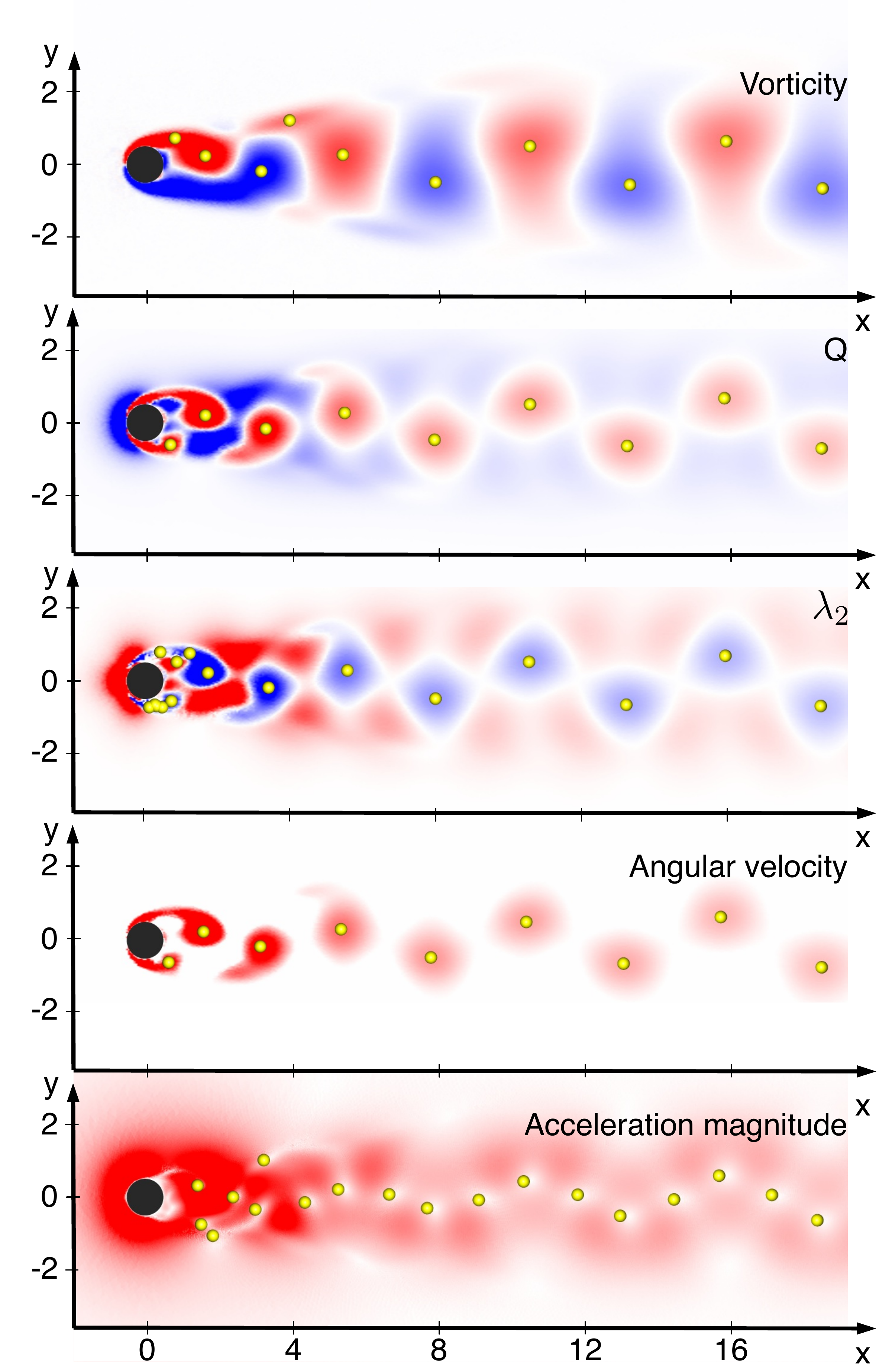}
\end{center}
\caption[X]{Visualization of a cylinder wake snapshot.
Five vorticity-related quantities 
are depicted by color maps (red: positive values, blue: negative, gray: zero):
(1) vorticity;
(2) Okubo-Weiss parameter; 
(3) $\lambda_2$; 
(4) absolute value of the imaginary part of the eigenvalues of the velocity Jacobian -- corresponds
to the angular velocity; 
(5) material acceleration magnitude. 
The yellow spheres depict the extremal points typically used as features for the respective
quantity.}
\label{fig:wake:5quantitites}
\end{figure}
%-----------------------------------------------------------------------
%----figure--------------------------------------------------------------
\begin{figure}[htb!]
%\vspace{10mm}
\begin{center}
\includegraphics[width=0.49\textwidth]{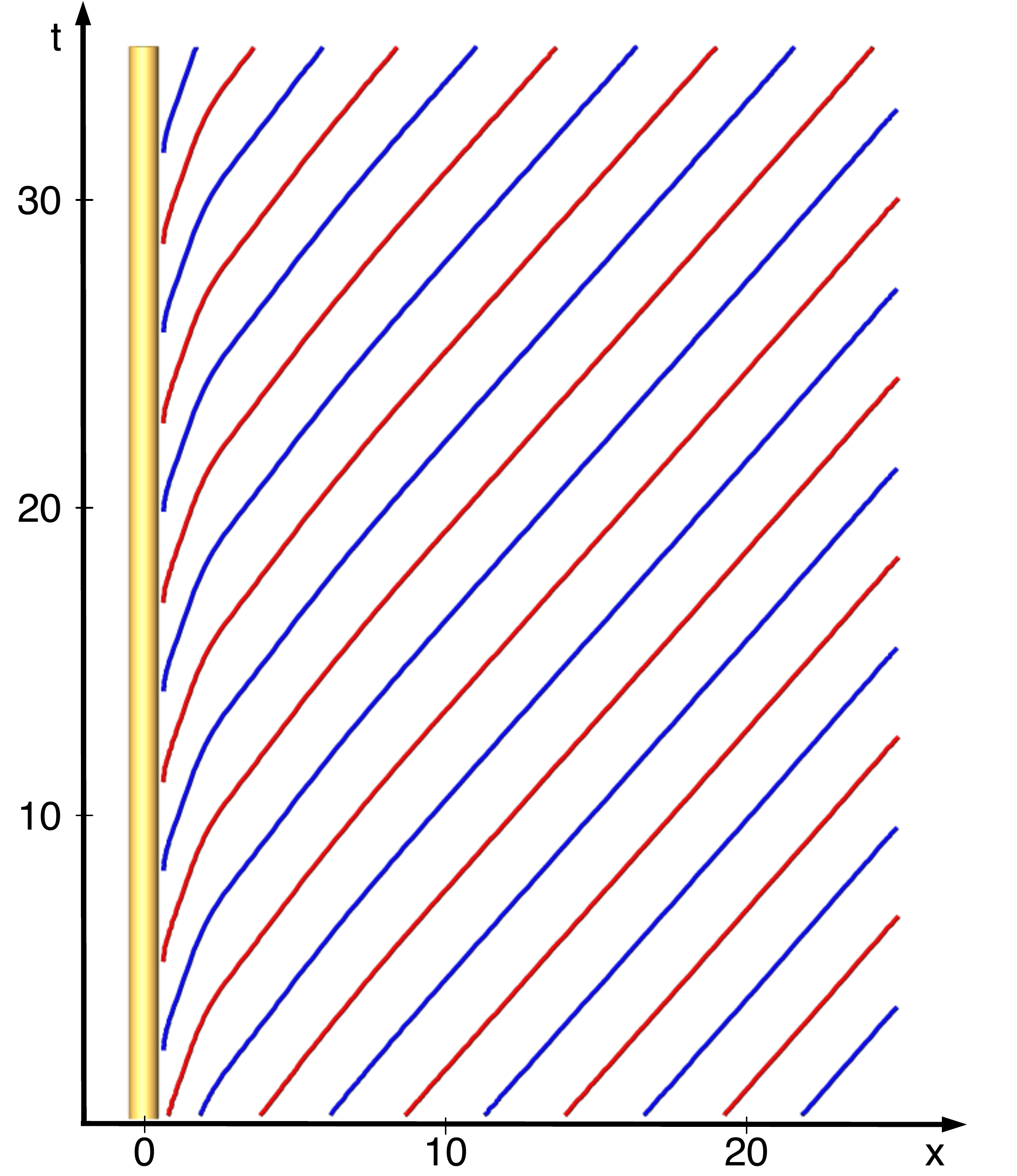}
\end{center}
\caption[X]{Tracked vortices of the cylinder wake in an $x$-$t$-view. 
Red (blue) marks positive (negative) rotation the vortices.}
\label{fig:wake:xt}
\end{figure}
%-----------------------------------------------------------------------
%----figure--------------------------------------------------------------
\begin{figure}[htb!]
%\vspace{10mm}
\begin{center}
\includegraphics[width=0.49\textwidth]{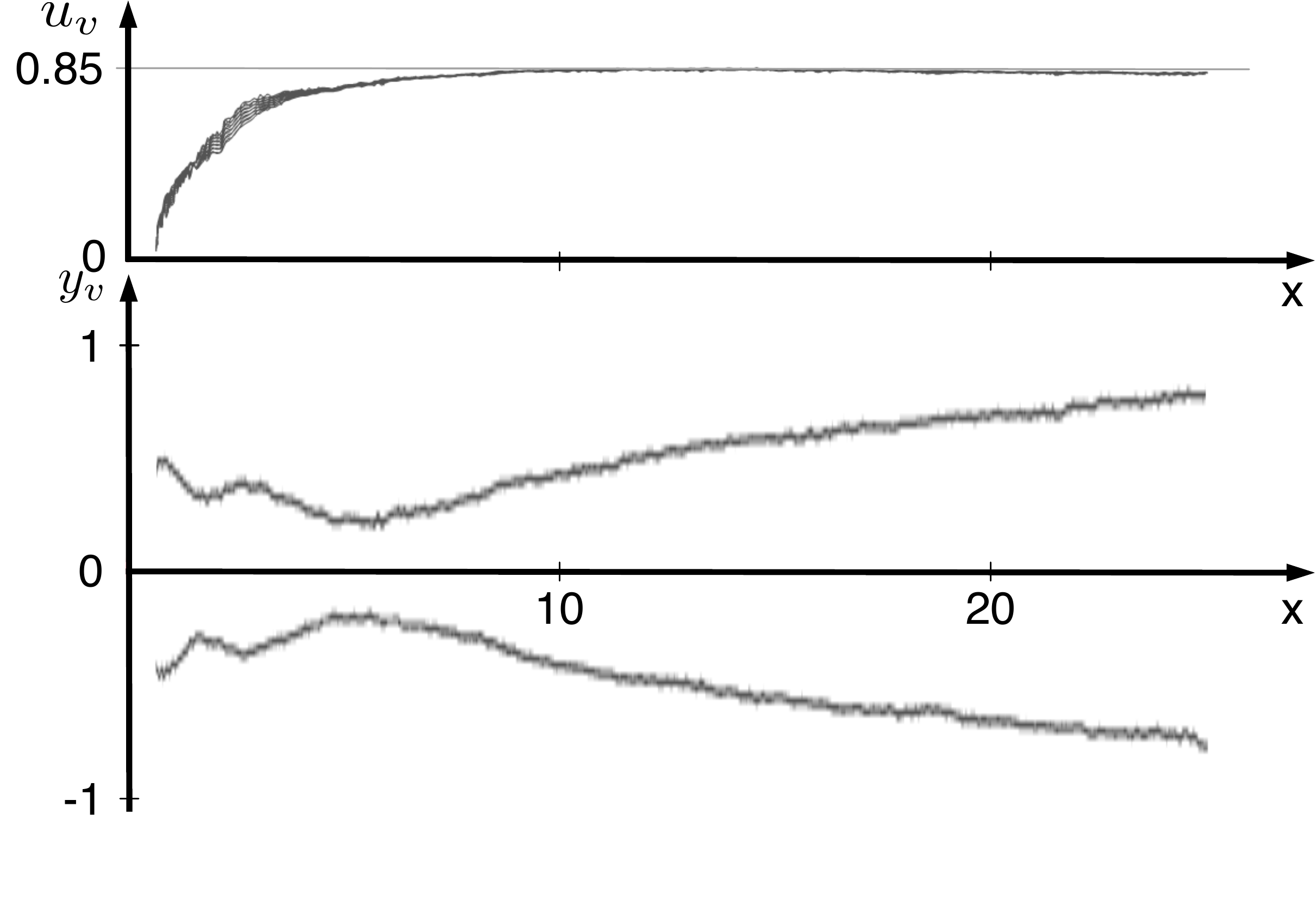}
\end{center}
\caption[X]{Plots of the streamwise velocity component $u_v$~(top)
and the transverse displacement $y_v$~(bottom) along
tracked vortices.
Note that each figure contains the history of many vortex evolutions
from roll-up to convection out of the domain.
Hence, several lines can be seen in each curve.  }
\label{fig:wake:qx}
\end{figure}
%-----------------------------------------------------------------------
Three free shear flows are investigated:
the cylinder wake (Sec.\ \ref{toc:wake}),
the mixing layer  (Sec.\ \ref{toc:mixinglayer}), and
the planar jet  (Sec.\ \ref{toc:jet}).
These configurations represent different levels of spatiotemporal complexity
from the periodic wake 
to the broadband dynamics and vortex pairing of the mixing layer and jet.
The first two flows share a pronounced uniform far-wake convection velocity,
while the jet structures move slower with streamwise distance.

Focus is placed on the vortex skeleton 
as identified by the acceleration feature points.
Pars pro toto, 
we perform a statistical analysis for the wake (Sec.\ \ref{toc:wake}),
investigate the vortex merging of the mixing layer  (Sec.\ \ref{toc:mixinglayer}), 
and employ the persistence-filter of LEPs for the jet  (Sec.\ \ref{toc:jet}).

All flows are described in a Cartesian coordinate system $\mathbf{x}=(x,y)$,
of which the abscissa $x$ points in streamwise direction
and $y$ in transversal direction.
The origin is located in the source of the shear flow,
i.e.\ 
center of the cylinder for the wake,
center of the inflow for the mixing layer
and center of the orifice for the jet.
The velocity $\mathbf{u}=(u,v)$ is expressed in the same system,
$u$ and $v$ being the $x$- and $y$-components of the velocity.
The time is denoted by $t$, the pressure by $p$ and the material acceleration by $\mathbf{a}$.
All quantities are non-dimensionalized with a characteristic length-scale $L$,
a characteristic velocity $U$ and the density of the fluid $\rho$.
$L$ denotes the cylinder diameter for the wake,
the vorticity thickness for the mixing layer,
and the width of the origin for the planar jet.
$U$ corresponds to the oncoming flow for the wake,
to the velocity of the upper (faster) stream for the mixing layer,
and to the maximum velocity at the orifice for the jet.

%----subsection---------------------------------------------------------
\subsection{Cylinder wake}
\label{toc:wake}
Starting point is a benchmark problem
of the data visualization community:
periodic vortex shedding behind a circular cylinder.
The Reynolds number is set to $100$,
which is well above the critical value for vortex shedding at $47$ 
\cite{Zebib1987jem,Jackson1987jfm}
and well below the critical value for transition-related instabilities
around $180$ \cite{Zhang1995pf,Williamson1996arfm}.
The flow is simulated with a finite-element method solver
with third-order accuracy in space and time, like in \cite{Noack2003jfm}.
The rectangular computational domain $(x,y) \in [-10,30] \times [-15,15]$
without the disk $K_{1/2}(0)$ for the cylinder 
is discretized by 277,576 triangular elements.
The numerical time step for implicit time integration is $0.1$, 
which also corresponds to the sampling frequency for the snapshots.

Figure\ \ref{fig:wake:5quantitites}  shows five vorticity related quantities of a cylinder wake snapshot.
The vorticity field depicts the separating shear-layers rolling up 
in a staggered array of alternating vortices.
The yellow balls mark the extrema, 
revealing the known fact that the ratio between 
the transverse of vortex displacement and the wavelength slightly 
increases downstream with vortex diffusion.
The second subfigure shows the Okubo-Weiss parameter 
$Q = \Vert \mathbf{S}^- \Vert^2 - \Vert \mathbf{S}^+ \Vert^2$
marking the maxima with balls.
This parameter employs the velocity Jacobian $\nabla \mathbf{u}$
and compares the norm of the symmetric shear tensor 
$\mathbf{S}^+\!=\!\frac{1}{2}\left[\nabla \mathbf{u} + \left( \nabla \mathbf{u} \right)^t \right]$
and with the norm of the antisymmetric one 
$\mathbf{S}^-\!=\!\frac{1}{2} \left[\nabla \mathbf{u} - \left( \nabla \mathbf{u} \right)^t \right]$.
In the center of a radially symmetric vortex,
$Q= \Vert \mathbf{\omega} \Vert^2  > 0$,
since $\Vert \mathbf{S}^+ \Vert$  vanishes and 
$\Vert\mathbf{S}^- \Vert$ becomes the norm of the vorticity $\Vert \mathbf{\omega} \Vert$.
At a saddle point $Q=  - \Vert \mathbf{S}^+ \Vert^2 < 0$.
Hence, maxima of $Q$ can be associated with vortex centers and minima with saddles.
The third subfigure shows $\lambda_2$.
Its extrema are marked by balls and indicate vortex centers.
$Q$ and $\lambda_2$ are generally considered to provide synonymous information.
The absolute value of the imaginary part of the Jacobian $\nabla \mathbf{u}$
is shown in the fourth subfigure. 
This quantity characterizes the angular frequency of revolution of a neighboring particles.
Hence, its maxima marked by yellow balls indicate vortex centers.
Finally, the magnitude of the material acceleration field is depicted.
The minima (zeros) mark both vortex centers and saddles, 
i.e.\ twice as many points in the vortex street.
These two features are distinguished based on the velocity Jacobian:
two positive eigenvalues  of the velocity Jacobian are associated with a saddle,
a complex conjugate pair with a vortex.

In the vortex street, 
all five vortex criteria provide nearly identical locations.
In the boundary-layer and in the near-wake of the cylinder
there are pronounced differences.
Mathematically, 
the vortex criteria rely on quite different formulae.
They cannot be expected to exactly coincide 
except for pronounced flow features, e.g.\ axis-symmetric vortices.
In addition, the cylinder boundary introduces 
a singular line $\mathbf{u} = \mathbf{0}$, thus amplifying the differences between the vortex criteria. 

Figure \ref{fig:wake:xt}
shows the spatial-temporal vortex evolution, based on the tracked acceleration feature points.
In the far-wake, a uniformly convecting von K\'arm\'an vortex street is observed.
In the near-wake, the convection speed is significantly slower.
This aspect is highlighted in Fig.\ \ref{fig:wake:qx} (first subfigure).
The streamwise velocity of each vortex $u_v$ 
is a monotonically increasing function from  0.03 to about 0.85. 
The asymptote corresponds to the literature value \cite{Williamson1996arfm}.
%----figure--------------------------------------------------------------
\begin{figure*}
%\vspace{10mm}
\begin{center}
\includegraphics[width=0.99\linewidth]{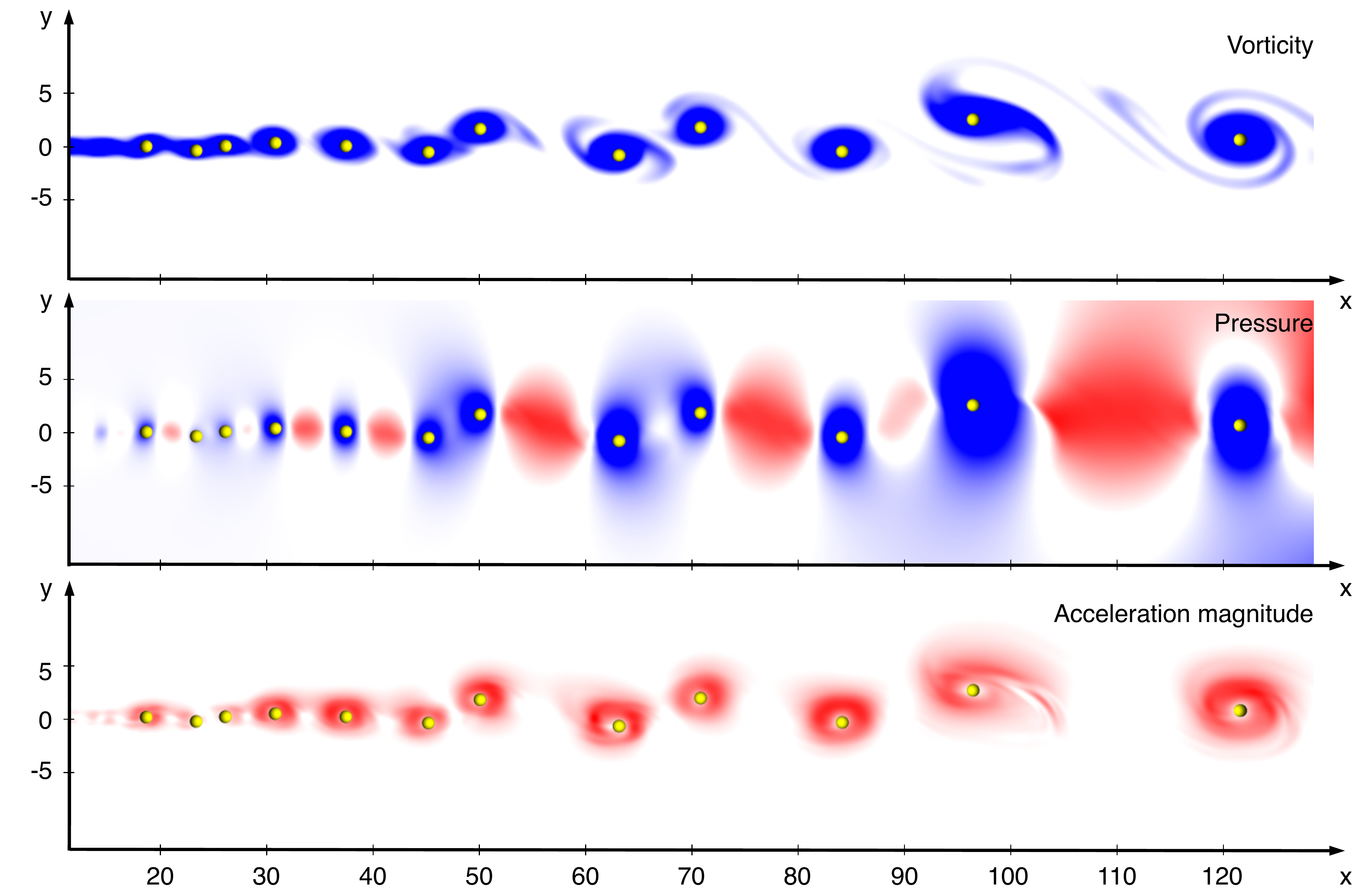}
\end{center}
\caption[X]{Visualization of a mixing layer snapshot.
Comparison of vorticity (top), pressure (middle) and acceleration (bottom). 
The color scheme is blue (red) for negative (positive) values.
The yellow spheres represent pronounced vortex acceleration feature points. }
\label{fig:mixinglayer:3quantities}
\end{figure*}
%-----------------------------------------------------------------------
The transverse spreading of the vortex street,
noted in Fig.\ \ref{fig:wake:5quantitites},
is quantified in the following subfigure with the transverse location $y_v$.

It should be noted that tracked acceleration feature points
can be seen as markers of coherent structures.
The acceleration-based framework provides a convenient means
for determining convection velocities and evolution of spatial extensions.
The following investigations of the mixing layer and the jet flow emphasize this aspect.
  
%----subsection---------------------------------------------------------

\subsection{Mixing layer}
\label{toc:mixinglayer}
The second investigated shear flow is a mixing-layer with
a velocity ratio between upper and lower stream of 3:1,
following earlier investigations of the authors
\cite{Comte1998ejmb,Noack2004swing,Noack2005jfm}.
The inflow is described by a $\tanh$ profile
with a stochastic perturbation.
And the Reynolds number based on maximum velocity 
and vorticity thickness is 500.
The flow is computed with a compact finite-difference scheme
of 6th order accuracy in space 
and 3rd oder accuracy in time.
The computational domain $(x,y) \in [0,140] \times [-28,28]$
is discretized on a $960\times384$ grid.
The sampling time for the employed snapshots is $\Delta t=0.05$
corresponding 1/10 of the computational time step.

In contrast to the space- and time-periodic Stuart solution,
the mixing layer generally shows several vortex pairing events.
In Fig.\ \ref{fig:mixinglayer:3quantities},
the distance between vortex acceleration feature points (marked by balls) 
are seen to increase in streamwise direction as result of vortex merging.
Furthermore, the locations of the acceleration feature points
nicely correlate with the local maxima of the vorticity (top),
the local minima of the pressure (middle) and the local minima
of the magnitude of the material acceleration (bottom).
The correlation between vorticity maxima and pressure minima
in free shear flows is well documented in the literature.
The correlation between pressure and acceleration magnitude minima
may be inferred from the non-dimensionalized Euler equation $\mathbf{a} = -\nabla p$,
governing the predominantly inviscid dynamics of the mixing layer
(see Sec.\ \ref{toc:method}).
A pressure minimum (or maximum)
implies $\nabla p =0$ and thus $\mathbf{a}=0$.

\begin{figure}
%\vspace{10mm}
\begin{center}
\includegraphics[width=0.8\columnwidth]{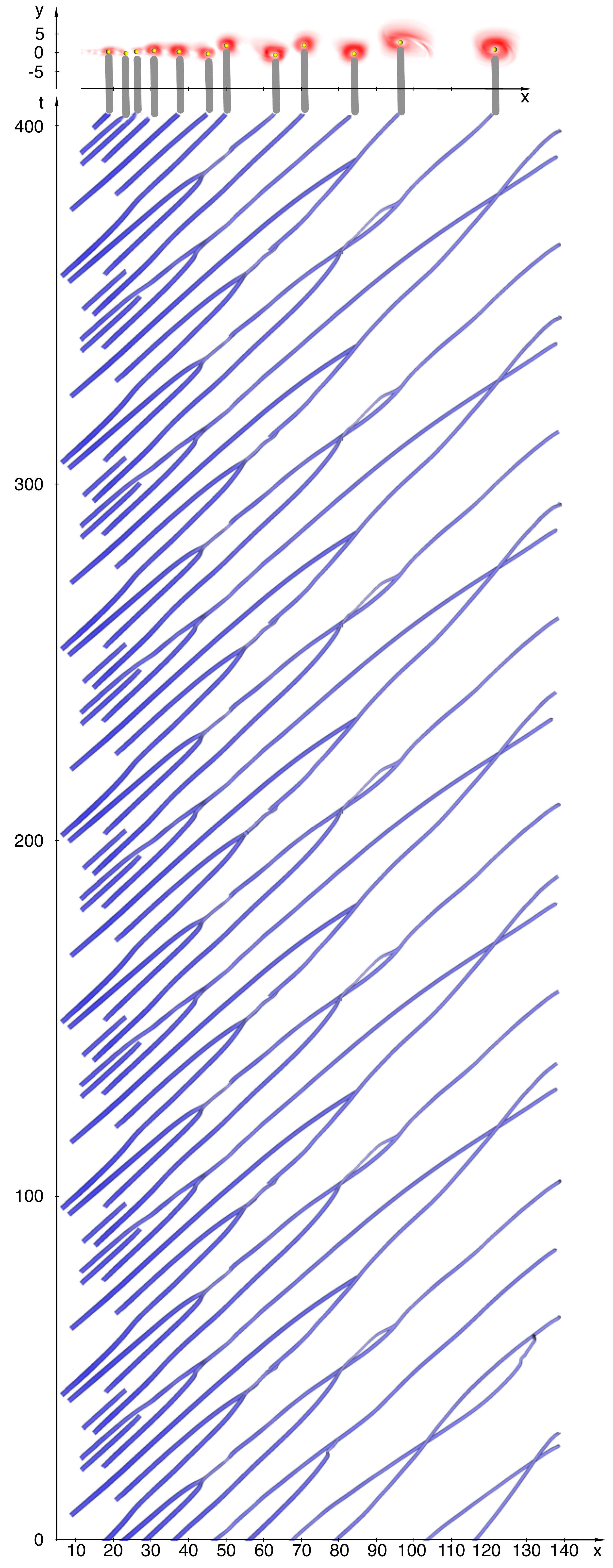}
\end{center}
\caption[X]{Spatiotemporal evolution of vortices in the mixing-layer.
The top part of the figure shows the acceleration magnitude field and LEPs at the final time considered.
The bottom figure marks the tracked acceleration feature points over approximately 5 downwash times.
Numerous vortex merging events can be identified. The size and coloring of the vortex skeleton is
determined by vorticity -- more intense blue corresponds to lower vorticity. Note that the vorticity is negative everywhere.}
\label{fig:mixinglayer:xt}
\end{figure}

The vortex merging events are shown in Fig.~\ref{fig:mixinglayer:xt}.
%----figure--------------------------------------------------------------
%-----------------------------------------------------------------------
Upstream, many Kelvin-Helmholtz vortices are formed.
In streamwise direction numerous merging events can be identified,
approximately 2 successive vortex mergers in the domain shown. 
Not all crossing of $x,t$-curves mark mergers 
since vortex pairs may rotate around their center
before eventual merging.
The figure strongly suggests a nearly constant streamwise convection velocity,
as expected from literature results
and contrary to the cylinder wake dynamics.

%----subsection---------------------------------------------------------
\subsection{Planar jet}
\label{toc:jet}
Finally, the spatiotemporal evolution of the planar jet is investigated.
Like the mixing layer, 
the jet shows a number of vortex mergers
leading to a reduction of the characteristic frequency.
As additional complexity, the convection velocity is not constant
but decreases in streamwise direction.

All quantities are normalized with the jet width $D_j$ 
and maximum jet velocity $U_j$. 
The flow is a weakly compressible isothermal 2D jet 
with a Mach number of $M\!a_j=0.1$ 
and a Reynolds number of $Re_j=D_j U_j/ \nu_{\infty}=500$.
The inflow velocity profile is given 
by a hyperbolic tangent profile like in \cite{Freund2001jfm}:
\begin{displaymath}
  u(r)=U_{\infty} + \frac{(U_j-U_{\infty})}{2} \left[ 1 - \tanh \left[ b \left( \frac{r}{r_0}- \frac{r_0}{r}\right) \right ] \right ].
\end{displaymath}   
Here, 
a uniform 1\% co-flow $U_{\infty} = 0.01 U_j$  
is added to avoid vortices with arbitrarily long residence time in the computational domain.
The slope of the $\tanh$ profile is characterized
by $b= {r_o} / {4\delta_{\theta}}$ and 
the momentum thickness of the shear layer is $\delta_{\theta}=0.05r_o$.
The initial mean temperature was calculated with the Crocco-Busemann relation, 
and the mean initial pressure  was constant.

The natural transition to unsteadiness is promoted 
by adding disturbances in a region in the early jet development 
near the inflow boundary $x_o=-0.5$ : 
\begin{equation}
  v(x,y)=v(x,y)+\alpha U_c e^{-\frac{(x-x_o)^2}{\lambda_x^2}}(f_1(y)+f_2(y))
\end{equation}
Here, 
\begin{equation}
  f_1(y)=\epsilon_1 e^{-\frac{(y-y_1)^2}{\lambda_y^2}}, \qquad 
  f_2(y)=\epsilon_2 e^{-\frac{(y-y_2)^2}{\lambda_y^2}},
\end{equation}
where $U_c=0.5$, $\alpha=0.008$, 
$y_1=0.5$, $y_2=-0.5$, $\lambda_x=0.1$, $\lambda_y=0.1$ and 
 $-1 \leq \epsilon_1,\epsilon_2 \leq 1 $ are random numbers.

The flow is defined in a rectangular domain $(x,y) \in [0,20] \times [-7,7]$.
The adjacent sponge zone extends to $[-1.5,25] \times [-10,10]$.
%\textcolor{magenta}{\bfseries WARNING: THE VISUALIZATION EXTENDS INTO THE SPONGE LAYER!}
The whole domain is discretized on a non-uniform Cartesian with $2\,449$ points
in $x$-direction and 598 points in $y$-direction.
The compressible Navier-Stokes equation is solved by mean of a (2,4) 
conservative finite-difference scheme based on MacCormack's 
predictor-corrector method \cite{Gottlieb1976mc} with block-decomposition and MPI parallelization. 
The system may be closed by the thermodynamic relations for an ideal gas. Details of the equations,
boundary conditions and solver can be inferred from \cite{Cavalieri2011jsv}.

\begin{figure*}
%\vspace{10mm}
\begin{center}
\includegraphics[width=0.99\textwidth]{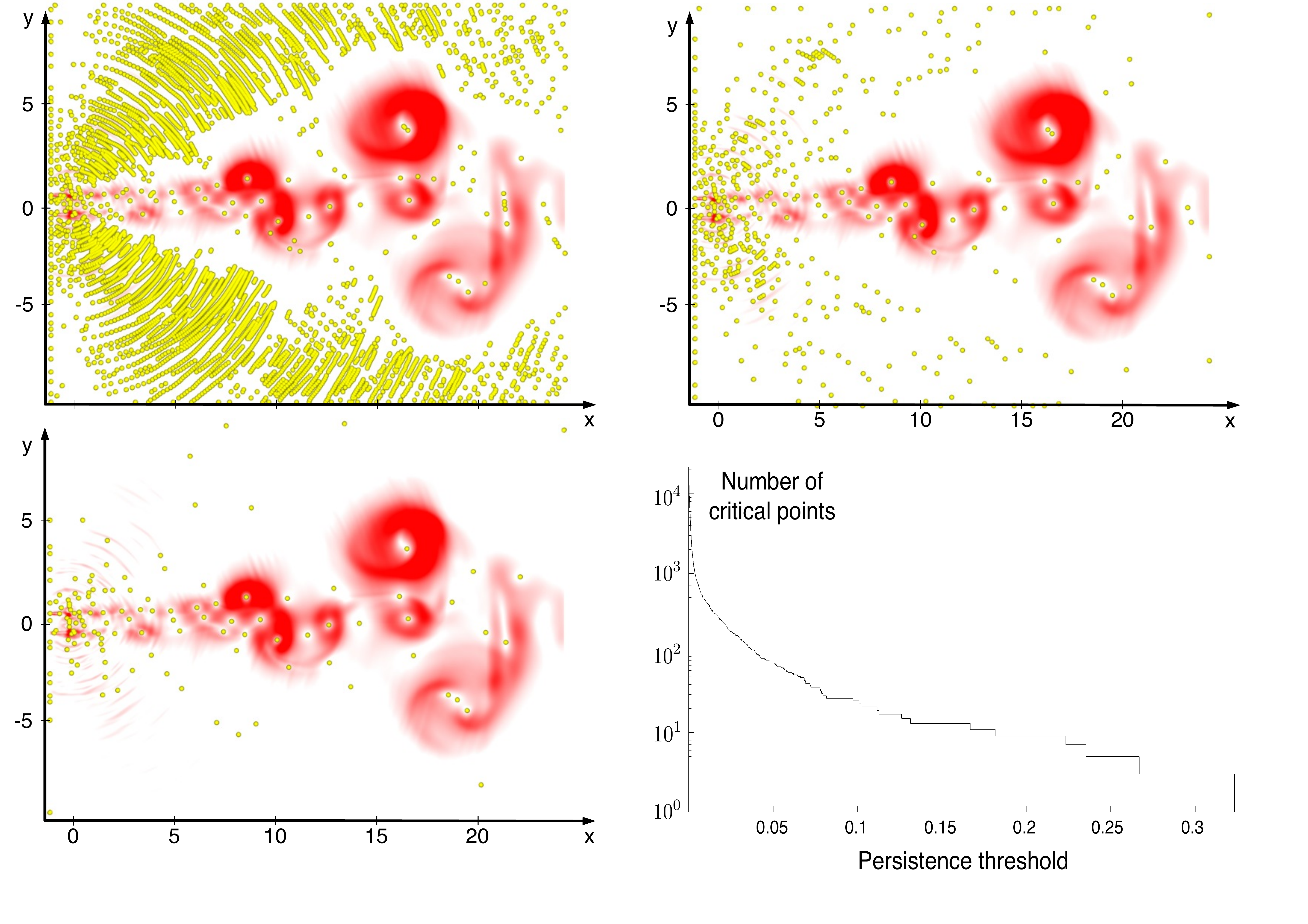}
\caption[X]{
Persistence-based visualization of a jet snapshot.
\emph{Top left to bottom left}: Visualization of the snapshot for persistence threshold levels
of $0\%$ (top left), $0.5\%$ (top right) and $2\%$ (bottom left) of the maximum.
The color field depicts the acceleration magnitude 
-- with a color map that ranges  from white (zero) to red (positive). 
The yellow balls represent acceleration feature points filtered by their persistent homology
with respect to the specified threshold levels.\\
\emph{Bottom right}: Persistence distribution. The number of critical points after persistence-based filtering
is plotted against
the persistence threshold level.
}
\label{fig:jet:persistence}
\end{center}
\end{figure*}

A stochastic inflow perturbation 
gives rise to small acoustic waves.
These small acoustic perturbations 
provide an excellent test-case
demonstrating the need and performance 
of persistence-based filtering.
Figure \ref{fig:jet:persistence} depicts a jet snapshot.
%----figure--------------------------------------------------------------
%------------------------------------------------------------------------
%----figure--------------------------------------------------------------
\begin{figure}
%\vspace{10mm}
\begin{center}
\includegraphics[width=0.56\columnwidth]{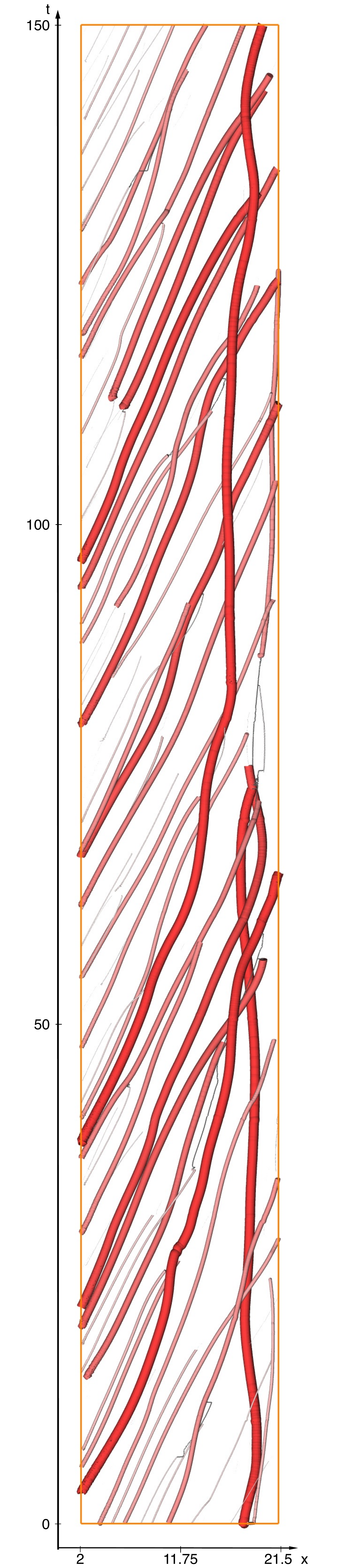}
\end{center}
\caption[X]{Spatiotemporal evolution of the vortex skeleton of the jet.
The size and coloring of the vortex lines are determined by our spatiotemporal importance measure. 
The links between the individual vortices are shown as white gray lines. }
\label{fig:jet:xt}
\end{figure}
%-----------------------------------------------------------------------
Most of the acceleration feature points are associated low-amplitude sound waves 
from the random inlet perturbation (top figure).
These acceleration feature points may be filtered out, ignoring those with low persistence (middle and bottom).
The bottom figure shows only features 
associated with incompressible dynamics.

\begin{figure*}
%\vspace{10mm}
\begin{center}
\includegraphics[width=0.98\textwidth]{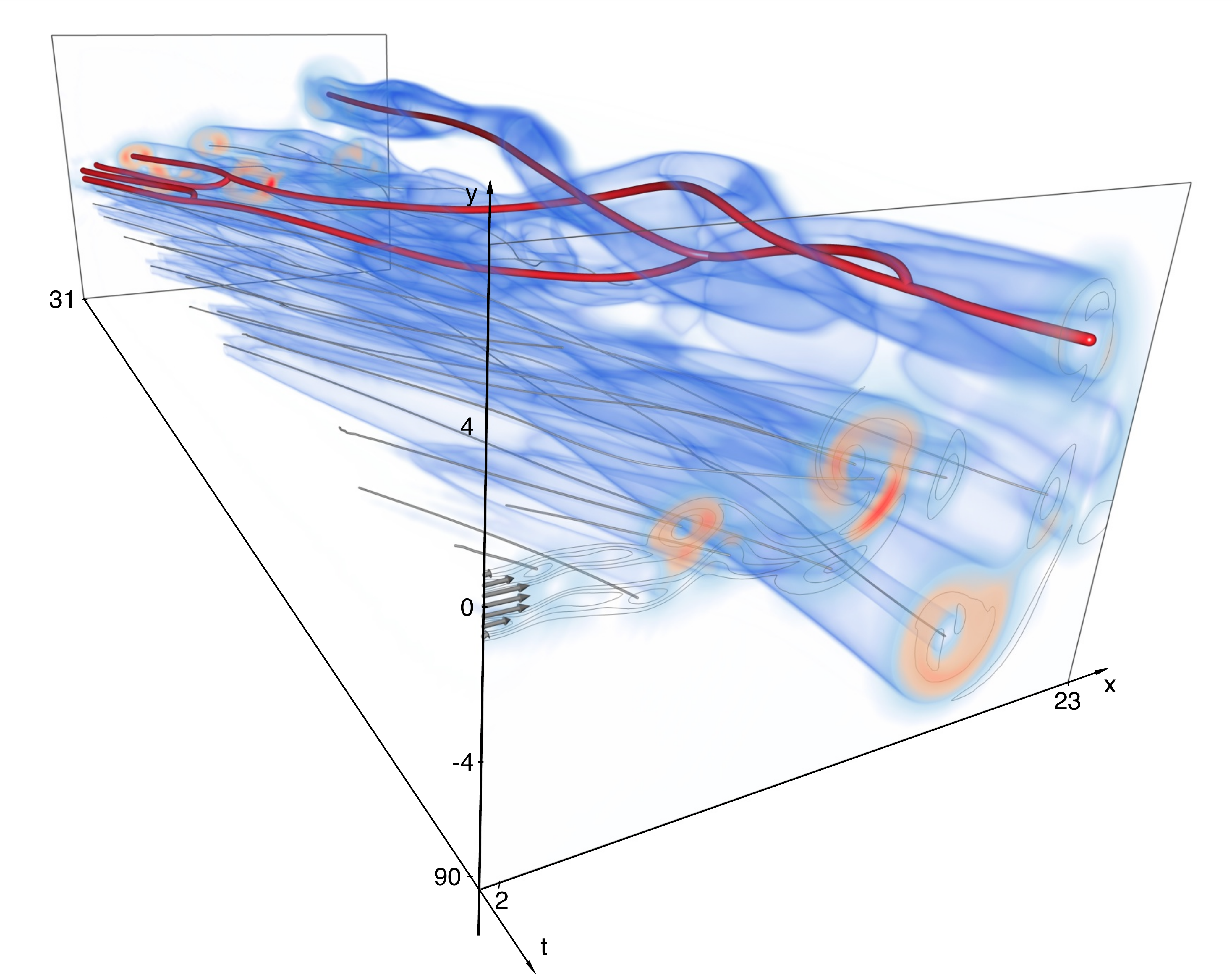}
\end{center}
\caption[X]{Close-up view of the vortex skeleton of the jet flow. The gray lines 
depict the extracted and filtered vortex cores. A few lines are visually highlighted
by red coloring; they show a pronounced vortex merging event and the origin of the merged vortices.
The acceleration is visualized by the blue volume
rendering and the color coding in the front and back plane. For comparison, iso-lines
of the vorticity are added to the front plane. }
\label{fig:jet:merge}
\end{figure*}

The spatiotemporal evolution of the vortex skeleton of the jet
is visualized in Fig.\ \ref{fig:jet:xt}
in a similar manner as the wake (Fig.\ \ref{fig:wake:xt}) 
and the mixing-layer (Fig.\ \ref{fig:mixinglayer:xt}).
Clearly, vortex merging events and a streamwise decreasing convection velocity
can be identified. 
In particular, some strong vortices remain for a long time near the exit.
A three-dimensional close-up view is shown in Fig.~\ref{fig:jet:merge}. 

%----figure--------------------------------------------------------------
%-----------------------------------------------------------------------%!TEX root = ./Kasten20111015pf.tex
%----section------------------------------------------------------------
\section{Conclusion and outlook}
\label{toc:conclusions}

We have proposed a novel feature extraction strategy for unsteady 2D flows.
This strategy departs in important aspects from 
topology extraction of the instantaneous velocity field,
starting from the velocity zeros.
Instead of the velocity, 
the material acceleration field is analyzed,
following \cite{Goto2006pf}.
Secondly, instead of acceleration zeros, 
the minima of the acceleration magnitude are identified. 
Thirdly, the acceleration feature points are tracked in time.
Finally, a mathematically rigorous spatiotemporal hierarchy of the tracked minima is defined.

The acceleration feature points define topological elements
of an unsteady flow with a number of discriminating features:
\begin{enumerate}
%-------------------
\item For steady flows, 
the acceleration feature points are a natural generalization of the critical points of vector field topology. 
Each critical point is an acceleration feature point.
This implication may not hold generally in the other direction.
%-------------------
\item A critical point of a steady flow field 
remains an acceleration feature point
in any inertial frame of reference.
In other words, 
the acceleration feature points
cannot vanish or be distorted 
by a uniform convection of a 'frozen' flow field (Taylor hypothesis).
%-------------------
\item The acceleration feature points are independent of the inertial frame of reference,
i.e.\ they are Galilean-invariant.
This property is a trivial consequence of the material acceleration field as observable.
%-------------------
%\item The fact that a LEP is based on the minimum of acceleration magnitude
%as opposed to its zero, implies robustness 
%with respect to a slowly accelerating frame of reference,
%e.g.\ one moving with the coherent structure.
%-------------------
\item The concept of acceleration feature points is parameter-free. 
No integration windows, nor threshold criteria, etc.\ are needed.
Note that the persistence level is not a free parameter,
as it defines a feature hierarchy.
%-------------------
\item Acceleration is correlated with pressure by neglecting the viscous term.
Suppose the pressure field has a minimum (in a vortex)
or maximum (near a saddle point).
Then, the pressure gradient vanishes
and the Euler equation yields a vanishing material acceleration
(implying trivially also a magnitude minimum).
%-------------------
\item The persistence measure introduces a rigorous hierarchy
of acceleration feature points based on spatial characteristics,
without the need of temporal filtering.
%-------------------
\item With the tracking of the acceleration feature points, 
the temporally integrated persistence
emphasizes long-lived structures.
\end{enumerate}
In short, identification of acceleration feature points naturally generalizes identification of critical points and exhibits new desirable or even necessary properties for a meaningful flow analysis.

Our framework follows Vassilicos' group \cite{Goto2006pf} 
in employing the Galilean-invariant material acceleration field 
as opposed to the velocity field.
However, Vassilicos determines the zeros of this field,
while our acceleration feature points are based on the more general notion of magnitude minima. 
This enables a robust, computationally inexpensive, derivative-free feature extraction 
--- capable of coping with large noise levels in the data. 
Furthermore, using minima instead of zeros allows for a natural extension to three-dimensional flows.
The concept of acceleration feature points follows Haller in the search of a Lagrangian Galilean invariant definition of saddles \cite{Haller2001physd} and vortices \cite{Haller2005jfm},
but provides a simple aggregate definition for both features.
The hierarchy of the acceleration feature points can be determined from a single snapshot, i.e.,  
no back-and-forward integration of fluid particles is required.

The framework has been applied to three free shear flows:
periodic vortex shedding of a cylinder wake,
a mixing layer with a small range of dominant frequencies,
and a planar jet with broadband dynamics.
In all cases, 
the acceleration feature points are cleanly distilled from the numerical data and they enable additional insights.
For the wake flow, vortex-based statistics are possible,
e.g., for determining the streamwise convection velocity.
For the mixing layer, vortex merging events are specified in time and space.
And for the jet, persistence is used to separate between aeroacoustic
and hydrodynamic equilibrium points.

In  the numerical analyses, only vortices have been considered.
Here, vortices are acceleration feature points with imaginary eigenvalues of the velocity Jacobian.
Analogously, saddles can be defined as acceleration feature points with real eigenvalues of this Jacobian matrix.
Thus, the concept of acceleration feature points represents a unifying framework for the main generic features of 2D flows.
Furthermore, it offers a computationally inexpensive alternative to the concept of the finite-time Lyapunov exponent \cite{Haller2001physd}.
We actively pursue a 3D generalization of the proposed feature extraction.

The Galilean-invariant generalization
of critical points has no obvious analogues to connectors in topology.
The very concept of connectors as boundary between different particles
looses its meaning in a time-dependent flow field.

As outlook, 
3D acceleration feature points (or lines and surfaces) 
with associated importance hierarchy and tracking technique
may pave the path to future kinematic analyses of complex turbulent flows.
A technique that extracts a hierarchy of 
the physically important points (and lines) of complex data 
will become invaluable as a first step in analysis.
The required visualization methods are already correspondingly mature.

%--- Acknowledgments --------------------------------------------------------------------------
\vspace*{1cm}
\section*{Acknowledgments}
The authors acknowledge funding of the German Research Foundation (DFG)
via the Collaborative Research Center (SFB 557) 
``Control of Complex Turbulent Shear Flows'' 
and the Emmy Noether Program.
Further funding was provided by the Zuse Institute Berlin (ZIB),
the DFG-CNRS research group 
``Noise Generation in Turbulent Flows'' (2003--2010),
the Chaire d'Excellence
'Closed-loop control of turbulent shear flows 
using reduced-order models' (TUCOROM)
of the French Agence Nationale de la Recherche (ANR), 
and the European Social Fund (ESF App. No. 100098251).
We thank the Ambrosys Ltd.\ Society for Complex Systems Management
and the Bernd R. Noack Cybernetics Foundation for additional support.
A part of this work was performed using HPC resources from GENCI-[CCRT/CINES/IDRIS] supported by the Grant 2011-[x2011020912].
We appreciate valuable stimulating discussions
with William K.\ George, Michael Schlegel, Gilead Tadmor, Vassilis Theofilis,
and Christos Vassilicos
as well as the local TUCOROM team:
Jean-Paul Bonnet,
Laurent Cordier,
Jo\"el Delville,
Peter Jordan, and
Andreas Spohn.
%We are grateful for  hardware and software support at Institut PPRIME by Alexandre Morel.{\color{red}TODO: Vielleicht noch nicht? :) :}
%Last but not least, we thank the referees for their thoughtful and helpful suggestions.
The figures have been created with Amira, a system for advanced visual data analysis (\texttt{http://amira.zib.de}).

%----bibliography--------------------------------------------------------------------------------
\bibliographystyle{unsrt}
\bibliography{arXiv.bib}

\begin{thebibliography}{10}

\bibitem{Lighthill1963book}
M.~J. Lighthill.
\newblock {\em Attachment and Separation in Three Dimensional Flow}.
\newblock Oxford University Press, Oxford, 1st edition, 1963.

\bibitem{Tobak1982arfm}
M.\ Tobak and D.~J. Peake.
\newblock Topology of three-dimensional separated flows.
\newblock {\em Ann.\ Rev.\ Fluid Mech.}, 14:61--85, 1982.

\bibitem{Perry1987arfm}
A.~E.\ Perry and M.~S. Chong.
\newblock A description of eddying motions and flow patterns using
  critical-point concepts.
\newblock {\em Ann.\ Rev.\ Fluid Mech.}, 19:125--155, 1987.

\bibitem{Rodriguez2010jfm}
D.\ Rodriguez and V.~Theofilis.
\newblock Structural changes of laminar separation bubbles induced by global
  linear instability.
\newblock {\em J.\ Fluid Mech.}, 655:280--305, 2010.

\bibitem{Rodriguez2011tcfd}
D.\ Rodriguez and V.~Theofilis.
\newblock On the birth of stall cells on airfoils.
\newblock {\em Theor.\ Comput.\ Fluid Dyn.}, 25:105--117, 2011.

\bibitem{Wang2006jfm}
L.\ Wang and N.~Peters.
\newblock The length-scale distribution function of the distance between
  extremal points in passive scalar turbulence.
\newblock {\em J.\ Fluid Mech.}, 554:457--475, 2006.

\bibitem{Wang2008jfm}
L.\ Wang and N.~Peters.
\newblock Length-scale distribution functions and conditional means for various
  fields in turbulence.
\newblock {\em J.\ Fluid Mech.}, 608:113--138, 2008.

\bibitem{Goto2006pf}
S.\ Goto and J.~C. Vassilicos.
\newblock Self-similar clustering of inertial particles and zero-acceleration
  points in fully developed two-dimensional turbulence.
\newblock {\em Phys.\ Fluids}, 18(11):115103--1..10, 2006.

\bibitem{Edelsbrunner2008}
H.~Edelsbrunner and J.~Harer.
\newblock Persistent homology --- a survey.
\newblock In J.~E. Goodman, J.~Pach, and R.~Pollack, editors, {\em Surveys on
  Discrete and Computational Geometry: Twenty Years Later}, volume 458, pages
  257--282. AMS Bookstore, 2008.

\bibitem{Reininghaus2010a}
J.~Reininghaus, D.~G{\"u}nther, I.~Hotz, S.~Prohaska, and H.-C. Hege.
\newblock {TADD}: A computational framework for data analysis using discrete
  {M}orse theory.
\newblock In {\em Proc. ICMS 2010}, 2010.

\bibitem{Reininghaus2011}
J.~Reininghaus, J.~Kasten, T.~Weinkauf, and I.~Hotz.
\newblock Efficient computation of combinatorial feature flow fields.
\newblock {\em IEEE Trans. Vis. Comput. Graph.}, to appear, 2012.

\bibitem{Stuart1967jfm}
J.T. Stuart.
\newblock On finite amplitude oscillations in laminar mixing layers.
\newblock {\em J.\ Fluid Mech.}, 29:417--440, 1967.

\bibitem{Cabral:1993}
B.~Cabral and L.~C. Leedom.
\newblock Imaging vector fields using line integral convolution.
\newblock In {\em Proceedings of the 20th Annual Conference on Computer
  Graphics and Interactive Techniques}, SIGGRAPH '93, pages 263--270, New York,
  NY, USA, 1993. ACM.

\bibitem{Stalling:1995}
D.~Stalling and H.-C. Hege.
\newblock Fast and resolution independent line integral convolution.
\newblock In {\em Proceedings of the 22nd Annual Conference on Computer
  Graphics and Interactive Techniques}, SIGGRAPH '95, pages 249--256, New York,
  NY, USA, 1995. ACM.

\bibitem{Wu2005axial}
J.Z. Wu, A.K. Xiong, and Y.T. Yang.
\newblock Axial stretching and vortex definition.
\newblock {\em Phys. Fluids}, 17:038108, 2005.

\bibitem{taylor1938spectrum}
G.~I. Taylor.
\newblock The spectrum of turbulence.
\newblock {\em Proceedings of the Royal Society of London. Series
  A-Mathematical and Physical Sciences}, 164(919):476--490, 1938.

\bibitem{Haller2005jfm}
G.~Haller.
\newblock An objective definition of a vortex.
\newblock {\em J.\ Fluid Mech.}, 525:1--26, 2005.

\bibitem{Panton1984book}
R.W. Panton.
\newblock {\em Incompressible {F}low}.
\newblock John Wiley \& Sons, New York, etc., 1984.

\bibitem{Basdevan1994pd}
Claude Basdevant and Thierry Philipovitch.
\newblock On the validity of the ``{W}eiss criterion'' in two-dimensional
  turbulence.
\newblock {\em Phys. D}, 73(1-2):17--30, May 1994.

\bibitem{Forman1998b}
R.~Forman.
\newblock Morse theory for cell-complexes.
\newblock {\em Advances in Mathematics}, 134(1):90--145, 1998.

\bibitem{robins10}
V.~Robins, P.~Wood, and A.~Sheppard.
\newblock Theory and algorithms for constructing discrete {M}orse complexes
  from grayscale digital images.
\newblock {\em IEEE Transactions on Pattern Analysis and Machine Intelligence},
  33(8):1646--1658, 2011.

\bibitem{Kasten2012b}
J.~Kasten, I.~Hotz, B.~R. Noack, and H.-C. Hege.
\newblock Vortex merge graphs in two-dimensional unsteady flow fields.
\newblock In {\em Proceedings Joint EG - IEEE TVCG Symposium on Visualization},
  2012.

\bibitem{Zebib1987jem}
A.~Zebib.
\newblock Stability of viscous flow past a circular cylinder.
\newblock {\em J.\ Engr.\ Math.}, 21:155--165, 1987.

\bibitem{Jackson1987jfm}
C.P. Jackson.
\newblock A finite-element study of the onset of vortex shedding in flow past
  variously shaped bodies.
\newblock {\em J.\ Fluid Mech.}, 182:23--45, 1987.

\bibitem{Zhang1995pf}
H.-Q.\ Zhang, U.\ Fey, B.~R.\ Noack, M.\ K\"onig, and H.~Eckelmann.
\newblock On the transition of the cylinder wake.
\newblock {\em Phys.\ Fluids}, 7(4):779--795, 1995.

\bibitem{Williamson1996arfm}
C.H.K. Williamson.
\newblock Vortex dynamics in the cylinder wake.
\newblock {\em Annu.\ Rev.\ Fluid Mech.}, 28:477--539, 1996.

\bibitem{Noack2003jfm}
B.~R.\ Noack, K.\ Afanasiev, M.\ Morzy\'nski, G.\ Tadmor, and F.~Thiele.
\newblock A hierarchy of low-dimensional models for the transient and
  post-transient cylinder wake.
\newblock {\em J.\ Fluid Mech.}, 497:335--363, 2003.

\bibitem{Comte1998ejmb}
P.\ Comte, J.H.\ Silvestrini, and P.~B\'egou.
\newblock Streamwise vortices in {L}arge-{E}ddy {S}imulations of mixing layer.
\newblock {\em Eur.\ J.\ Mech. B}, 17:615--637, 1998.

\bibitem{Noack2004swing}
B.~R.\ Noack, I.\ Pelivan, G.\ Tadmor, M.~Morzy\'nski, and P.~Comte.
\newblock Robust low-dimensional {G}alerkin models of natural and actuated
  flows.
\newblock In {\em {\rm W.\ Schr\"oder \& P.\ Tr\"oltzsch,} Fourth Aeroacoustics
  Workshop}. Institut f\"ur Akustik und Sprachkommunikation, Technische
  Universit\"at Dresden, 2004.

\bibitem{Noack2005jfm}
B.~R.\ Noack, P.\ Papas, and P.~A. Monkewitz.
\newblock The need for a pressure-term representation in empirical {G}alerkin
  models of incompressible shear flows.
\newblock {\em J.\ Fluid Mech.}, 523:339--365, 2005.

\bibitem{Freund2001jfm}
J.B. Freund.
\newblock Noise sources in a low {R}eynolds number turbulent jet at {M}ach 0.9.
\newblock {\em J.\ Fluid Mech.}, 438:277--305, 2001.

\bibitem{Gottlieb1976mc}
D.~Gottlieb and E.~Turkel.
\newblock Dissipative two-four method for time dependent problems.
\newblock {\em Math. of Comp.}, 30(136):703--723, 1976.

\bibitem{Cavalieri2011jsv}
A.~V.~G.\ Cavalieri, G.\ Daviller, P.\ Comte, P.\ Jordan, G.\ Tadmor, and Y.\
  Gervais.
\newblock Using large eddy simulation to explore sound-source mechanisms in
  jets.
\newblock {\em J.\ Sound Vibr.}, 330(17):4098--4113, 2011.

\bibitem{Haller2001physd}
G.~Haller.
\newblock Distinguished material surfaces and coherent structures in {3D} fluid
  flows.
\newblock {\em Phys. D}, 149(4):248--277, 2001.

\end{thebibliography}
%\begin{thebibliography}{10}
%\end{thebibliography}

\end{document}